\documentclass[a4paper,11pt]{article}
\usepackage{amsfonts,amsmath}
\usepackage{amssymb}
\usepackage[dvips]{graphicx}
\usepackage{amscd}
\usepackage{xypic}
\usepackage[mathscr]{eucal}
\usepackage{amsthm}
\newtheorem{theorem}{Theorem}[section]
\newtheorem{definition}[theorem]{Definition}
\newtheorem{proposition}[theorem]{Proposition}
\newtheorem{lemma}[theorem]{Lemma}
\newtheorem{corollary}[theorem]{Corollary}
\newtheorem{remark}[theorem]{Remark}

\renewcommand{\theequation}{\thesection.\arabic{equation}}
\def\ds{\displaystyle}
\def\H{\mathcal H}
\def\endproof{\ \hfill\hbox{\vbox{\hrule\hbox{\vrule
height5pt\kern5pt\vrule height5pt}\hrule}}\par\medskip\rm}
\evensidemargin \oddsidemargin 
\date{}
\title{\textbf{Geometric approach  to the Hamilton-Jacobi equation and global parametrices for the Schr\"odinger propagator}}
\author{Sandro Graffi and Lorenzo Zanelli   \\
\\
{\small Dipartimento di Matematica, Universit\`a di Bologna,} \\
{\small Piazza di Porta S. Donato 5,
40126 Bologna, Italy.} \\
{\small (graffi@dm.unibo.it, zanelli@dm.unibo.it)}   }

\begin{document}
\baselineskip=18pt
\maketitle
\begin{abstract} \par\noindent
We construct a family of  Fourier Integral Operators, defined for arbitrary large times, representing a global parametrix for the Schr\"odinger propagator when the potential is quadratic at infinity.  This construction is based on the geometric approach to the corresponding Hamilton-Jacobi equation and thus sidesteps the problem of the caustics generated by the classical flow. Moreover, a detailed study of the real phase function allows us to recover a WKB semiclassical approximation which necessarily involves the multivaluedness of the graph of the Hamiltonian flow past the caustics. 
\end{abstract}
\tableofcontents
\newpage

\section{Introduction and statement of the results}
\renewcommand{\theequation}{\thesection.\arabic{equation}}
\setcounter{equation}{0}%
Let us consider the  initial value problem for the Schr\"odinger equation:
\begin{equation}
\label{eqSch} \left\{
\begin{array}{l}{\ds
i \hbar \partial_t \psi(t,x) =   - \frac{\hbar^2}{2m}  \Delta \psi(t,x) + V(x) \psi(t,x),}
\\
\\
\psi(0,x) = \varphi(x).
\end{array}
\right.
\end{equation}
where the potential  $V \in C^\infty (\Bbb R^n;\Bbb R)$ is assumed  quadratic at infinity.
 In this case it is well known that the operator $H$ in $L^2(\Bbb R^n)$ defined by the maximal action of $\ds - \frac{\hbar^2}{2m}  \Delta  + V(x) $ is self-adjoint. Hence the Cauchy problem (\ref{eqSch}) considered in $L^2(\Bbb R^n)$ admits the unique global solution $\ds \psi(t,x)=e^{-iHt/\hbar}\varphi(x)$, $\forall\,t\in{\Bbb R}$, $\forall\,\varphi\in L^2(\Bbb R^n)$. \par 
Under the present conditions a parametrix of the propagator under the form of a semiclassical Fourier integral operator (WKB representation)  has been constructed long ago by Chazarain \cite{Chz} (for related results by the same technique see also \cite{DF}, \cite{Ki}; for recent related work see  \cite{M-Y}, \cite{K1}, \cite{K2}).   The occurrence of caustics of the Hamilton-Jacobi equation makes this construction  local in time; the solution at an arbitrary time $T>0$ requires  multiple compositions of the local representations. A  global parametrix for the propagator has been constructed through the method of  complex valued phase functions (as in  \cite{KS},  \cite{L-S}), with related complex transport coefficients. A particularly convenient choice of the complex phase function (the Herman-Kluk representation) has been isolated in the chemical physics literature long ago (\cite{H-K}). Its validity has been recently  proved in \cite{Sw-R} and \cite{Rob2}). The relation between the above approaches and the underlying classical flow  is however less direct than the standard WKB approximation in which the phase function solves the Hamilton-Jacobi equation.

In this paper we study the problem through the geometric approach to the Hamilton-Jacobi equation (see e.g.\cite{C-Z1}, \cite{Sik86}).  In Theorem 1.1 a parametrix is obtained for the propagator $\ds U(t):=e^{i Ht/\hbar}
$  valid for $t \in [0,T]$, $0<T<\infty$,  under the form of a family of semiclassical {\it global}  Fourier Integral Operators (FIO), which extend to continuous operators in $L^2({\Bbb R}^n)$. The corresponding phase function is {\it real} and generates the graph of the  flow of the Hamiltonian $ \ds \mathcal{H} = \frac{p^2}{2m} + V(x)$. This technique not only yields
globality in time, but also helps to obtain a unified view of Fujiwara's as well as
Chazarain's approaches on one side, and of the Laptev-Sigal one on the other. In Theorem 1.2 we prove that a  WKB construction is still valid, necessarily multivalued because of the caustics. 

We assume:
\begin{eqnarray}
\label{H1}
&&
 V(x)= \langle L x , x \rangle + V_0 (x), \quad L\in GL(n), \quad det \ L \neq 0; 
\\
&&
\label{H2}
 V_0 \in C^\infty (\Bbb R^n), \quad| \partial_x^{\alpha} V_0 (x) | \le C_0.
\end{eqnarray}
Then the main result of the paper is:
\begin{theorem}  Let (\ref{H1}) and (\ref{H2}) be fulfilled. Let $0<T<\infty$ and $\varphi \in \mathcal{S}(\Bbb R^n)$.  Then:
\begin{equation}
\label{rep-00}
\psi(t,x) = (2\pi\hbar)^{-n}  \sum_{j=0}^\infty \int_{\Bbb R^n} \int_{\Bbb R^n} \int_{\Bbb R^k}   e^{\frac{i}{\hbar} ( S( t,x,\eta,\theta)  - \langle y , \eta \rangle )   }   \hbar^j b_{j} ( t,x,\eta,\theta) \ d\theta  \ d\eta  \  \varphi(y) \ dy + O(\hbar^\infty)  
\end{equation}
Here:
\begin{equation} \label{upperbound}
k >   C \  T^4 \;  \sup_{|\alpha|+|\beta| \ge 2}\;\sup_{(x,p)\in{\Bbb R}^{2n}}    |\partial^\alpha_x \partial^\beta_p {\mathcal H} (x,p) |^2
\end{equation}
for some $C>0$.  
Moreover the following assertions hold:
\par\noindent
(1) 
$S$ generates the graph $\Lambda_t$ of the Hamiltonian flow $\phi_{\mathcal H}^t : T^\star \Bbb R^{n} \rightarrow T^\star \Bbb R^{n}$  $\forall\,t\in [0,T]$:
\begin{eqnarray}
\Lambda_t &:=& \left\{  (y,\eta;x,p) \in T^\star \Bbb R^n \times T^\star \Bbb R^n \ | \ (x,p) = \phi_\H^t (y,\eta)     \right\}
\nonumber\\
&=& \left\{ (y,\eta;x,p) \in T^\star \Bbb R^n \times T^\star \Bbb R^n \ | \  p = \nabla_x S, \; y =  \nabla_\eta S, \;  0 =  \nabla_\theta S \right\}  
\label{gen-lagr}
\end{eqnarray}
(2)
 $S \in C^\infty ([0,T] \times \Bbb R^{2n} \times \Bbb R^k ; \Bbb R)$ and  has the expression:
\begin{eqnarray}
\label{F-gen01}
S &=&  \langle x , \eta \rangle   - \frac{t}{2m} \eta^2 -  t \langle L x, x \rangle   +   \langle Q(t)  \theta  ,   \theta   \rangle 
+ \langle v(t,x,\eta) , \theta + f(t,x,\theta)  \rangle + \langle \nu(t,x,\eta,\theta) , \theta  \rangle 
\nonumber\\
&+&  g (t,x,\eta,\theta).
\end{eqnarray}
Here $f(t,x,\theta): [0,T] \times {\Bbb R}^n \times {\Bbb R}^k \to {\Bbb R}^k$, $\nu(t,x,\eta,\theta): [0,T]\times{\Bbb R}^n\times{\Bbb R}^n\times {\Bbb R}^k \to {\Bbb R}^k$,  $g(t,x,\eta,\theta): [0,T] \times {\Bbb R}^n\times{\Bbb R}^n\times {\Bbb R}^k\to {\Bbb R} $ and  $C_{\alpha \beta \sigma}(T)>0$ are such that
$$
\sup_{[0,T]\times{\Bbb R}^{2n+k}} [|\partial_x^\alpha \partial_\theta^\sigma f (t,x,\eta,\theta)| +|\partial_x^\alpha \partial_\eta^\beta \partial_\theta^\sigma g (t,x,\eta,\theta)|+|\partial_x^\alpha \partial_\eta^\beta \partial_\theta^\sigma \nu (t,x,\eta,\theta)|]\le C_{\alpha \beta \sigma}(T). 
$$ 
The function $(x,\eta)\mapsto v(t,x,\eta): {\Bbb R}^n\times{\Bbb R}^n\to {\Bbb R}$ is linear $\forall\,t\in {\Bbb R}$, and $t\mapsto Q(t): [0,T]\to GL(k)$ with $Q(0)=0$. 
\vskip 4pt\noindent
(2)
The transport  coefficients $b_j: j=0,\ldots$ are determined by the first order PDE:
\vskip 5pt\noindent
\begin{eqnarray}
&&
\label{eq-tr0} \left\{
\begin{array}{l}
\partial_t b_0 +  \frac{1}{m} \nabla_x S \ \nabla_x b_0  + \frac{1}{2m} \Delta_x S  \ b_{0}   (t,x,\eta,\theta) = \Theta_N , 
\\
\\
b_0  ( 0,x,\eta,\theta) =   \rho(\theta).
\end{array}
\right. \quad j=0
\\
&&
\label{trasp-i} \left\{
\begin{array}{l}
\displaystyle{ \partial_t b_j +  \frac{1}{m} \nabla_x S \ \nabla_x b_j  + \frac{1}{2m} \Delta_x S  \ b_{j}  -  \frac{i}{2m}  \Delta_x  b_{j-1}  = 0 ,}
\\
\\
b_j (0,x,\eta,\theta) = 0, \quad \rho(\theta)\in  \mathcal{S} (\Bbb R^k;\Bbb R^+),  \;\| \rho \|_{L^1} = 1.
\end{array}
\right.  \quad j \ge 1
\end{eqnarray}
Here $\Theta_N  \in C^\infty_b (\Bbb R^{2n+k}; \Bbb R)$ is arbitrary within the requirement:
\begin{eqnarray*}
&&
\Pi^\alpha \Theta_N \in C^\infty_b (\Bbb R^{2n+k}; \Bbb R), \;0\leq \alpha\leq N;
\\
&&
\Pi  \Theta_N := {\rm div}\, \left(  \Theta_N \frac{ \nabla_\theta S}{|\nabla_\theta S|^2} \right).
\end{eqnarray*}
(3) $\forall\,0 \le t \le T$, $0 \le T < + \infty$, the expansion (\ref{rep-00})  generates an $L^2$ parametrix of the propagator $\ds U(t)= e^{i Ht/\hbar}$: each term   is  a continuos FIO  on $\mathcal{S}(\Bbb R^n)$ denoted $B_j(t)$, $j=0,1,\ldots$, which admits a continuous extension to  $L^2 (\Bbb R^n)$, and:
\begin{equation}
\label{parametrice}
 e^{i Ht/\hbar}=\sum_{j=0}^\infty\,B_j(t)+O(\hbar^{\infty}).
\end{equation}
The notation $O(\hbar^\infty)$ means: 
$$
\| R_{N}(t) \|_{L^2\to L^2} \le C_{N}(T) \hbar^{N+1},  \forall\, N\ge0,\;\forall\,t\in[0,T], \quad R_{N}(t) := U(t) -  \sum_{j=0}^N B_j(t).
$$
 Moreover, the expansion (\ref{parametrice}) does not depend on $\rho$ provided $\|\rho\|_{L^1}=1$. Namely, if $\rho_1\neq \rho_2$:
 $$
 \sum_{j=0}^N\,B_j[\rho_1](t)-\sum_{j=0}^N\,B_j[\rho_2](t)=O(\hbar^{N+1}).
 $$
\end{theorem}
\vskip 4ptBy applying the stationary phase theorem to the oscillatory integral (\ref{rep-00}), the integration over the auxiliary parameters $\theta$ can be eliminated and  the  WKB approximation to the evolution operator is recovered, necessarily multivalued on account of the caustics.

\begin{theorem} 
Let $V(x) = \frac{1}{2}|x|^2 + V_0 (x)$ with $ \sup_{x \in \Bbb R^n} \| \nabla^2 V_0(x)\| < 1$; let $\widehat{\varphi}_\hbar (\eta)$ be the $\hbar$-Fourier transform of the initial datum $\varphi$. 
Then $\forall\,t\in [0,T]$, $t \neq (2\tau+1) \frac{\pi}{2}$, $\tau \in \mathbb{N}$, there exists a finite open partition  $\displaystyle \Bbb R^n \times \Bbb R^n = \bigcup_{\ell=1}^{\mathcal{N}} D_\ell$ such that the solution of  (\ref{eqSch}) can be represented as:
$$
\psi(t,x) = \int_{\Bbb R^n}  \widehat{U}_\hbar (t,x,\eta) \ \widehat{\varphi}_\hbar (\eta) \ d\eta , \quad 0\leq t\leq T,\quad t \neq (2\tau+1) \frac{\pi}{2}
$$
\begin{equation}
\label{wkb99}
\widehat{U}_\hbar (t,x,\eta) \Big|_{D_\ell} =   \sum_{\alpha=1}^\ell   \ e^{\frac{i}{\hbar} S_{\alpha}(t,x,\eta)    }  \   |{\rm det} \nabla^2_\theta S(t,x,\eta,\theta^\star_\alpha (t,x,\eta))|^{-\frac{1}{2}}  e^{\frac{i\pi}{4}\sigma_\alpha}   b_{\alpha,0} (t,x,\eta)   + O(\hbar)
\end{equation}
\begin{eqnarray}
S_\alpha  := S ( t,x,\eta,\theta^\star_\alpha (t,x,\eta)), \
b_{\alpha,0}  := b_{0} (t,x,\eta,\theta^\star_\alpha (t,x,\eta)), \  \sigma_\alpha := sgn \nabla^2_\theta S ( t,x,\eta,\theta^\star_\alpha (t,x,\eta)) 
\nonumber
\end{eqnarray}
where $\mathcal{N}$ is a $t-$dependent natural and:
\begin{enumerate}
\item[(i)]
On each $D_\ell$  the equation $0=\nabla_\theta S(t,x,\eta,\theta)$ has $\ell$ smooth solutions $\theta^\star_\alpha (t,x,\eta)$, $1\le \alpha \le \ell$.
\item[(ii)]  Any function $S_\alpha (t,x,\eta)$ solves  locally the Hamilton-Jacobi equation:
\begin{equation}
\nonumber
\frac{|\nabla_xS_\alpha |^2}{2m} (t,x,\eta)+V(x)+\partial_t S_\alpha (t,x,\eta) =0
\end{equation}
\item[(iii)] 
An explicit upper bound  on the $t$-dependent natural $\mathcal{N}$ is computed in (\ref{eq-loc-N}). 
\end{enumerate}
\end{theorem}

\noindent
{\bf Example} In the  harmonic oscillator case $V(x) = \frac{1}{2} x^2$ and the  phase function is exactly quadratic 
$
S(t,x,\eta,\theta) = \langle x  , \eta \rangle   - \frac{t}{2} (  \eta^2 +  x^2 )  + \langle v(t,x,\eta) , \theta \rangle +  \langle Q(t) \theta  , \theta \rangle
$ 
It admits  a unique smooth global critical point $\theta^\star(t,x,\eta)$ on $(x,\eta) \in \Bbb R^{2n}$ for $ t\in [0,T]$, $t \neq (2\tau+1) \frac{\pi}{2}$, $\tau \in \mathbb{N}$. Hence the series (\ref{wkb99}) reduces to just one term conciding with the well known Mehler formula:
\begin{eqnarray*}
\psi(t,x)  =   \int_{\Bbb R^n} e^{\frac{i}{\hbar\cos(t)}  \left( \langle x  , \eta \rangle - \frac{\sin(t)}{2} (\eta^2 + x^2)  \right)  }   \frac{1}{\cos(t)}  \hat{\varphi}_\hbar (\eta)     d\eta 
\end{eqnarray*}
{\bf Remarks}
\begin{enumerate}
\item 
The phase function is constructed  (Section 2) through  the Amann-Conley-Zehnder reduction technique of the  action functional (\cite{A-Z}, \cite{C-Z1},\cite{C2}).  Namely:
\begin{equation}
\label{fgen00}
S ( t,x,\eta,\theta)  =    \langle x , \eta \rangle  +  \int_0^t  [\gamma^p (s)  \dot{\gamma}^x (s)  -  H(\gamma^x (s), \eta +  \gamma^p(s))] \ ds \Big|_{\gamma (t,x,\theta)(\cdot)}
\end{equation}
where the  curves  $\Gamma(t,x,\theta)=(\gamma^x (t,x,\theta)(s), \gamma^p (t,x,\theta)(s))$ are parametrized as follows:
\begin{equation}
\Gamma(t,x,\theta):=\left\{
\begin{array}{l}
\displaystyle{\gamma^x (t,x,\theta)(s) = x  -   \int_s^t  \phi^x (t,x,\theta) (\tau) \ d\tau ,
\quad \phi^x  = \theta^x (\cdot) +  f^x  (t,x,\theta)(\cdot)}, 
\\
\\
\displaystyle{ \gamma^p (t,x,\theta)(s) =   \int_0^s  \phi^p (t,x,\theta) (\tau) \ d\tau ,
 \quad   \phi^p = \theta^p (\cdot)  +  f^p  (t,x,\theta)(\cdot)}
\end{array}
\right.
\label{set-curves}
\end{equation}
Here  $\theta \in  \mathbb{P}_M L^2 ([0,T];\Bbb R^{2n}) \simeq \Bbb R^k$ ($\mathbb{P}_M$ is the  finite dimensional Fourier orthogonal projector, $k= 2n(2M+1)$) so that the parameters $\theta$ can be identified with  the  finite Fourier components of the derivatives of the curves $\gamma$. (\ref{fgen00}) represents a global generating function if  $k$ fulfills the lower bound (\ref{upperbound}). In turn, the functions $(f^x, f^p) : [0,T] \times\Bbb R^n \times  \mathbb{P}_M L^2 \rightarrow  \mathbb{Q}_M L^2 \times  \mathbb{Q}_M L^2$ are determined by a fixed point functional equation, essentially the $\mathbb{Q}_M$ projection of the Hamilton equations (Section 2.3).\\
The parametrization (\ref{gen-lagr}) entails that $S$ is a smooth solution of the problem: 
\vskip 4pt\noindent
\begin{equation}
\label{eqHJw}
\left\{
\begin{array}{l} {\ds 
\frac{|\nabla_x  S |^2}{2m} (t,x,\eta,\theta) + V(x) + \partial_t  S  (t,x,\eta,\theta) = 0, }
\\
\\
S  ( 0,x,\eta,\theta) = \langle x , \eta \rangle; \quad  \nabla_\theta S ( t,x,\eta,\theta) = 0. 
\end{array}
\right.
\end{equation}
\item
Any function $S(t,x,\eta,\theta)$ solving (\ref{eqHJw}), i.e. the Hamilton-Jacobi equation under the stationarity constraint $\ds  \nabla_\theta S  = 0$, is the central object to determine the so called {\it geometrical solutions of the Hamilton-Jacobi equation} (see for example the recent works \cite{C2}, \cite{C4}).  Global generating functions are clearly not unique and this is due to the presence of the $\theta$-auxiliary parameters.  Uniqueness holds instead for the geometry of set of critical points:
$$
\Sigma_S :=\{(x,\eta,\theta )\in{\Bbb R}^{2n+k}\;|\;\nabla_\theta S ( t,x,\eta,\theta) = 0\} 
$$
which does not depend on $S$ because it is globally diffeomorphic to $\Lambda_t$; a detailed study of $\Sigma_S$ is done in Section 2. We  prove (Section 3) that symbols  coinciding on $\Sigma_S$ generate semiclassical Fourier Integral Operators differing only by terms $O(\hbar^\infty)$. This will allow us to select  symbols  in such a way to make essentially trivial the proof of the $L^2$ continuity of the associated operator. 
\item 
The symbol $b_0$ solving the geometrical version (\ref{eq-tr0}) of the transport equation is
\begin{equation}
\label{b0-def}
b_{0} (t,x,\eta,\theta)  =  \exp \left\{ - \frac{1}{2m} \int_0^t \Delta_x  S( \tau , \gamma^x (t,x,\theta)(\tau) ,\eta,\theta)  d\tau \right\}  \rho(\theta)
\end{equation}
If $T_2>T_1$, then  $k(T_2)>k(T_1)$  so that $\Gamma (T_1,x, \theta)\subset \Gamma(T_2,x,\theta)$. In the limit $T \rightarrow \infty$, $\theta \to \phi \in L^2 (\Bbb R^+; \Bbb R^{2n})$ and we get the simplified functional (still  well defined):  
\begin{equation}
\label{b0-def2}
b_{0} (t,x,\phi)  =  \exp \left\{  \frac{1}{2m} \int_0^t \Delta_x  V(  x  -   \int_\tau^t  \phi^x  (\lambda) \ d\lambda )  d\tau \right\}  \rho(\phi)
\end{equation}
This  corresponds to the zero-th order symbol of the Laptev-Sigal construction  \cite{L-S}: 
$$
v_0 (t,y,\eta) = \exp \left\{  \frac{1}{2m} \int_0^t \Delta_x V (  x^\tau(y,\eta)  )  d \tau \right\}.
$$ 
Namely,  the functional is the same, but is  evaluated on the classical curves (with initial conditions  $x^0(y,\eta)=y$, $p^0(y,\eta)=\eta$) instead of all the free curves used in (\ref{b0-def2}), with regularity $H^1$ and boundary condition $\gamma^x (t,x,\phi)(t)=x$. 
\item
 For potentials in the class (\ref{H1}) and $0 \le t \le T$ small enough no caustics develop, and there is a unique smooth solution $\theta^\star (t,x,\eta)$ for $(x,\eta)\in \Bbb R^{2n}$. The stationary phase theorem yields the $0$-th order approximation to the integral  (\ref{rep-00}):
\begin{eqnarray}
\label{ordinezero}
\widehat{U}_\hbar^{(0)} (t,x,\eta)=e^{\frac{i}{\hbar} S (t,x,\eta,\theta^\star )}    |{\rm det} \nabla^2_\theta S(t,x,\eta,\theta^\star )|^{-\frac{1}{2}} e^{\frac{i\pi}{4}\sigma}  b_{0} (t,x,\eta,\theta^\star)  
+
O(\hbar)
\end{eqnarray}
which coincide with the WKB semiclassical approximation. This fact suggests a relationship, at any order in $\hbar$, between the present construction and those of Chazarain \cite{Chz} and Fujiwara \cite{DF}.  This is the contents of Theorem \ref{th-corr86}.
\item
The first  three assertions of Theorem 1.2  represent the counterpart (in the $\eta$ variables) of a result of  Fujiwara  \cite{DF}, valid under the additional assumption  that  the number of classical curves connecting boundary data is finite. 
\end{enumerate}
{\small We thank Johannes Sj\"ostrand for suggesting us the formulation of Theorem 1.2, and Kenji Yajima for a critical reading of a first draft of this paper.}

\vskip 1cm\noindent
\section{Generating functions for the graph of the Hamiltonian flow}
\renewcommand{\theequation}{\thesection.\arabic{equation}}
\setcounter{equation}{0}%
\subsection{Lagrangian submanifolds and  global generating functions}
Adopting standard notations and terminology (see e.g.\cite{W1}), we denote by $\ds \omega=dp\wedge
dx=\sum_{i=1}^ndp_i\wedge dx^i$  the  2--form on
$T^\star\mathbb{R}^{n}$ that defines its natural symplectic
structure. As usual, a diffeomorphism
$\mathcal{C}:T^\star\mathbb{R}^{n} \rightarrow
T^\star\mathbb{R}^{n}$ is a {\it canonical transformation} if  the pull
back of the symplectic form is preserved, $\mathcal{C}^\star
\omega = \omega$.\\
We say that  $L\subset T^\star\mathbb{R}^{n}$ is a {\it Lagrangian
submanifold} if $\omega |_{L}=0$ and $\dim {L} =n=\frac{1}{2}
{\dim T^\star\mathbb R^n}$. In a natural way, a symplectic structure
$\bar\omega$ on  $T^\star\mathbb R^n\times T^\star\mathbb R^n \cong
T^\star\left(\mathbb R^n\times \mathbb R^n\right)$ is the twofold
pull--back of the standard symplectic 2--form on $T^\star\mathbb
R^n$ defined as $\bar
\omega:=pr_2^\star\omega-pr_1^\star\omega=dp_2\wedge dx_2-dp_1\wedge
dx_1$. Similarly, $\Lambda \subset T^\star\mathbb{R}^{n} \times
T^\star\mathbb{R}^{n}$ is called a Lagrangian submanifold of
$T^\star\mathbb{R}^{n} \times T^\star\mathbb{R}^{n}$ if $\bar
\omega |_{\Lambda} =0$ and $\mbox{\rm dim}(\Lambda)$ = 2n. \\
A Hamiltonian is a $C^2$-function $\H:
T^\star\mathbb{R}^{n} \rightarrow \mathbb{R}$ and its flow is the
one-parameter group of canonical transformations $\phi_\H^t : U
\subseteq T^\star\mathbb{R}^{n} \rightarrow
T^\star\mathbb{R}^{n}$ solving Hamilton's equations
$\dot{\gamma} = J \nabla \H (\gamma)$ ($J$ the unit symplectic matrix) with initial conditions
$\gamma(0)=(x_0,p_0) \in U$.\\
The  Hamilton-Helmholtz  functional:
\begin{eqnarray}
\label{action}
A [ (\gamma^x,\gamma^p) ] :=  \int_0^t  [\gamma^p (s)  \dot{\gamma}^x (s)  -  \H(\gamma^x (s),\gamma^p(s))] \ ds
\end{eqnarray}
is well defined and continuous on the path space $H^1 ([0,t];T^\star \Bbb R^n)$. The action functional:
\begin{eqnarray}
\mathcal{A} [\gamma^x] := \int_0^t \mathcal{L}(\gamma^x (s), \dot{\gamma}^x (s))  \ ds
\end{eqnarray}
is defined on  $H^1 ([0,t]; \Bbb R^n)$. In this paper we consider    $\ds \H= \frac{p^2}{2m} + V(x)$,  so  that the Legendre transform guarantees the corrispondence of the stationary curves of these two functionals.

\begin{definition}
\label{fgen-def}
A global generating  function for a Lagrangian submanifold $L \subset
T^{\star} \mathbb{R}^{n}$ is a $C^2$ function $S: \Bbb R^n\times\Bbb
R^k \rightarrow \Bbb R$ such that
\begin{itemize}
\item[$\diamond$]
$L= \left\{  (x, p ) \in T^{\star} \mathbb{R}^{n} | \,\,\, p =
\nabla_x S(x,\theta),  \, \, \, 0 = \nabla_\theta S(x,\theta) \   \right\} $,
\item[$\diamond$]
$
rank \left(  \nabla^2_{x\theta} S \  \nabla^2_{\theta \theta} S \right) \Big|_L  = {\rm max}. 
$
\end{itemize}
Similarly, a global generating  function for a Lagrangian submanifold
$\Lambda \subset T^{\star} \mathbb{R}^{n} \times T^{\star}
\mathbb{R}^{n}$ is a $C^2$ map $S : \Bbb R^n \times  \Bbb
R^n \times  \Bbb R^k \rightarrow \Bbb R$ such that
\begin{itemize}
\item[$\diamond$]
$\Lambda= \left\{  (x,p; y,\eta) \in T^{\star} \mathbb{R}^{n}  \times T^{\star} \mathbb{R}^{n}  | \,\,\, p =
\nabla_x S(x,\eta,\theta),  \,\,\, y =   \nabla_\eta S(x,\eta,\theta) ,    \, \, \, 0 = \nabla_\theta S \   \right\} $,
\item[$\diamond$]
$
rank \left(  \nabla^2_{x \theta} S  \  \nabla^2_{\eta \theta} S \ \nabla^2_{\theta \theta} S \right) \Big|_\Lambda  = {\rm max}. 
$
\end{itemize}
\end{definition}
It is important to remark that the following set:
\begin{equation}
\label{sigmaS}
\Sigma_S := \{ (x,\eta,\theta) \in \Bbb R^n \times \Bbb R^n \times \Bbb R^k \ | \ 0 = \nabla_\theta S(x,\eta,\theta)   \}
\end{equation}
is a submanifold of $\Bbb R^{2n+k}$ and it is diffeomorphic to $\Lambda$. \\
\indent We focus our attention on the graphs of a Hamiltonian flow $\phi_\H^t : T^\star \Bbb R^n  \rightarrow  T^\star \Bbb R^n$,  which correspond to a  family of Lagrangian submanifolds in $T^{\star} \mathbb{R}^{n}  \times T^{\star} \mathbb{R}^{n}$:
$$
\Lambda_t := \left\{  (y,\eta;x,p) \in T^\star \Bbb R^n \times T^\star \Bbb R^n \ | \ (x,p) = \phi_\H^t (y,\eta)     \right\}
$$ 
An important object in what follows is  the family of global generating functions:
$$
\Lambda_t = \left\{ (y,\eta;x,p) \in T^\star \Bbb R^n \times T^\star \Bbb R^n \ | \  p = \nabla_x S, \quad y =  \nabla_\eta S, \quad   0 =  \nabla_\theta S ( t,x,\eta,\theta)  \right\}  
$$
to be explicitly  constructed for arbitrarily large times in the next Section. As is known, this technical tool has been developed in the framework of symplectic geometry and variational analysis (see \cite{A-Z},  \cite{C-Z1}, \cite{Cha}, \cite{LSik}, \cite{V}, \cite{Sik86}, \cite{Sik}) to sidestep the locality in time generated by the occurrence of caustics.

\subsection{Generating function with infinitely many parameters}
In the following we mainly review the construction of a generating function with infinitely many parameters described in \cite{C2}.
We begin by the following simple result (see \cite{W1}):
\begin{lemma}
Let us consider the Hamilton-Helmholtz functional $A[\cdot]$ as in (\ref{action}). 
A curve $\gamma \in \Gamma^{(0)} := \{ \gamma \in H^1 ([0,t];T^\star \Bbb R^n)\ | \  \gamma^p (0)=0, \ \gamma^x (t)= x  \}$ satisfies Hamilton's equations with boundary conditions:
$$
\dot{\gamma} = J \nabla \H (\gamma), \quad  \gamma^p (0) = 0, \quad  \gamma^x (t) = x
$$  
if and only if the following stationarity condition of variational type holds:
$$
\frac{DA}{D\gamma} (\gamma) [v] = 0 \quad \forall v \in T \Gamma^{(0)} 
$$
\end{lemma}
\proof By computing the G\^ateaux derivative of the functional, we get: 
$$
\frac{DA}{D\gamma}(\gamma) [v] = \int_0^t \ [ \dot{\gamma} - J \nabla \H (\gamma)  ](s) v(s) \ ds +  \gamma^p (s) v^x (s) |_0^t  \quad \forall v \in T \Gamma^{(0)}.
$$
Now use the  boundary condition $\gamma^p (0) = 0$ and recall that for  $T \Gamma^{(0)}$ it must be $v^x (t)=0$. The result is proved.
\endproof
The above Lemma has an important consequence:  it allows us to introduce the notion of generating function with infinitely many parameters.\\
First of all, it is easy to observe that the set of curves 
\begin{eqnarray}
\label{set-gen}
\gamma(t,x,\phi)(s)  := \left(  x  -   \int_s^t  \phi^x (\tau) d\tau  ,
 \int_0^s \phi^p (\tau) d\tau  \right) \quad \quad  \phi \equiv (\phi^x,\phi^p)
\end{eqnarray}
gives a parametrization of the path space $\Gamma^{(0)}$ introduced in the previous lemma, namely:
$$
\Gamma^{(0)}(t,x,\phi) :=  \left\{ \gamma(t,x,\phi)(\cdot) \ | \ \phi \in L^2 ([0,T]; \Bbb R^{2n}) \right\}
$$
Second, we define the functional with {\it infinitely many parameters} specified by $\phi \in L^2 ([0,T]; \Bbb R^{2n})$ in the following way:
\begin{definition}
\begin{equation}
\label{fg-i2}
\mathcal{S} (t,x,\eta,\phi) :=    \langle x, \eta \rangle +   \int_0^t  [\gamma^p (s)  \dot{\gamma}^x (s)  -  H(\gamma^x (s), \eta + \gamma^p(s))] \ ds \Big|_{\gamma (\cdot) = \gamma(t,x,\phi)(\cdot)}
\end{equation} 
\end{definition}
\noindent
{\bf Remark}
\newline\noindent
Introducing the traslated curves  $\zeta = (\zeta^x,\zeta^p) := (\gamma^x, \eta + \gamma^p)$, it is easy to see that the functional (\ref{fg-i2}) admits the equivalent representation
$$
\mathcal{S} (t,x,\eta,\phi) =    \langle \zeta^x(0), \eta \rangle +   \int_0^t  \zeta^p (s)  \dot{\zeta}^x (s)  -  H(\zeta^x (s), \zeta^p(s)) \ ds \Big|_{\zeta (\cdot) = \zeta(t,x,\phi)(\cdot)}
$$
where now $\zeta^x (t)=x$ and $\zeta^p(0)=\eta$.\\

Let us now make our assumptions on  the Hamiltonian $\H$ more precise:
\begin{definition}
\begin{equation}
\label{H-def1}
\H(x,p) = \frac{p^2}{2m} + V(x) =  \frac{p^2}{2m} +  \langle L x , x \rangle + V_0(x)
\end{equation}
where $V_0 \in C^\infty (\Bbb R^n)$, $L\in GL(n)$,  and 
$$
| \partial_x^{\alpha} V_0 (x) | \le C_0,  
$$
\end{definition}

This allows us to look more closely at the structure of the generating function:
\begin{lemma}
The functional $\mathcal{S}$ admits the representation:
\begin{eqnarray}
\mathcal{S} (t,x,\eta,\phi)   &=&  \langle x , \eta \rangle   - \frac{t}{2m} \eta^2 - t \langle L x , x \rangle  +  \langle R(t) \phi , \phi  \rangle +  \langle v(t,x,\eta) , \phi \rangle + \sigma (t,x,\phi) .
\nonumber\\
\label{rep-S}
\end{eqnarray}
Here $v(t,x,\eta)$ has a linear dependence with respect to $(x,\eta)$ variables, and $\sigma (t,x,\phi)$ is bounded with respect to $(x,\phi)$.
\end{lemma}
\proof
It is easy to see that:
\begin{eqnarray}
\mathcal{S}   &=& \langle x , \eta \rangle +   \int_0^t  \left(   \int_0^s \phi^p (\tau) d\tau   \right)   \phi^x (s)   ds
\nonumber\\
&-&   \int_0^t  \frac{1}{2m} \left(\eta +  \int_0^s \phi^p (\tau) d\tau\right)^2  +    L \left(  x  -   \int_s^t  \phi^x (\tau) d\tau \right) ,  \left( x -  \int_s^t  \phi^x (\tau) d\tau \right)  ds
\nonumber\\
&-&   \int_0^t V_0 \left( x  -  \int_s^t  \phi^x (\tau) d\tau \right)  ds
\nonumber\\
&=&  \langle x , \eta \rangle   - \frac{t}{2m} \eta^2 - t \langle L x , x \rangle  +  \langle R(t) \phi , \phi  \rangle +  \langle v(t,x,\eta) , \phi \rangle + \sigma (t,x,\phi) 
\label{rep-S99}
\end{eqnarray}
where
\begin{eqnarray}
\langle R(t)\phi,\phi\rangle& :=&\int_0^t  \left(    \int_0^s \phi^p (\tau) d\tau     \phi^x (s)    -  \frac{1}{2m}  \left(      \int_0^s \phi^p (\tau) d\tau  \right)^2  
-    L   \int_s^t  \phi^x (\tau) d\tau      \int_s^t  \phi^x (\tau) d\tau   \right)   ds
\nonumber\\
\langle v(t,x,\eta) , \phi \rangle&:=&  \int_0^t \left(   - \frac{\eta}{m}  \int_0^s \phi^p (\tau) d\tau    +   2  L  x     \int_s^t  \phi^x (\tau) d\tau      \right) ds
\nonumber\\
\sigma (t,x,\phi)  &:=&  - \int_0^t   V_0 \left( x  -  \int_s^t  \phi^x (\tau) d\tau \right) \ ds
\end{eqnarray}
The boundedness of $\sigma$ is immediate:
$$
\sup_{(x,\phi) \in \Bbb R^n \times L^2 } |\sigma (t,x,\phi) | \le t \sup_{z \in \Bbb R^n} | V_0 (z) |
$$
Finally, we consider the orthonormal basis of $L^2$, $e_\alpha  (s)  = \frac{1}{\sqrt{T}} e^{ \frac{2\pi}{T} i \alpha s}$, ${\alpha \in \mathbb{Z}}$ and the corresponding Fourier expansion $\ds \phi(s) = \sum_{ \alpha \in \mathbb{Z}} \phi^{(\alpha)} e_\alpha (s)$. This entails the identification $\phi(s) \equiv \{ \phi^{(\alpha)}  \}_{ \alpha \in \mathbb{Z}} \in \ell^2$ under the usual norm $|\phi| = \sum_\alpha |\phi^{(\alpha)}|^2$ generated by the scalar product $\langle \psi , \phi \rangle = \sum_\alpha \psi^{(\alpha)} \phi^{(\alpha)}$.\\
\endproof

\begin{proposition}
\label{prop-sv0}
The graph of the Hamiltonian flow  
$$
\Lambda_t := \left\{  (y,\eta;x,p) \in T^\star \Bbb R^n \times T^\star \Bbb R^n \ | \ (x,p) = \phi_\H^t (y,\eta)     \right\}
$$ 
is generated by  $\mathcal{S}$:
$$
\Lambda_t = \left\{ (y,\eta;x,p) \in T^\star \Bbb R^n \times T^\star \Bbb R^n \ | \  p = \nabla_x \mathcal{S}, \quad y = \nabla_\eta \mathcal{S}, \quad   0=  \frac{ D \mathcal{S} }{D\phi} \right\}
$$
\end{proposition}
\proof
The first component of the stationarity equation $\ds 0= \frac{ D \mathcal{S} }{D\phi}$  reads:
$$
0  =  \frac{ D \mathcal{S} }{D\phi^p} (\phi) [v^p] =   \int_0^t  \left(    \int_0^s v^p (\tau) d\tau   \right)     \phi^x (s)  
-  \frac{1}{m} \left(  \eta +    \int_0^s \phi^p (\tau) d\tau  \right) \left(    \int_0^s v^p (\tau) d\tau  \right)     ds 
$$
for all $v^p \in L^2$. This is satisfied if and only if 
\begin{equation}
\label{eq-pf00} 
\phi^x (s) = \frac{1}{m} \left( \eta +  \int_0^s \phi^p (\tau) \ d\tau  \right)
\end{equation}
that is
\begin{equation}
\dot{\gamma}^x (t,x,\phi) (s) = \frac{1}{m}  \left( \eta +  \gamma^p (t,x,\phi)(s)  \right)
\end{equation}
On the other hand, the second equation reads:
$$
0 =  \frac{ D \mathcal{S} }{D\phi^x} (\phi) [v^x] =   \int_0^t  \left(    \int_0^s \phi^p (\tau) d\tau   \right)    v^x (s)   +  \nabla V \left( x -  \int_s^t \phi^x (\tau) d \tau \right)  \int_s^t v^x (\tau) d \tau \ ds
$$
for all $v^x \in L^2$. Integrating by parts, we get
\begin{eqnarray}
0 &=&   \int_0^t  \phi^p (s) ds    \int_0^t   v^x (\tau)    d\tau   -   \int_0^t   \phi^p (s)  \int_0^s  v^x (\tau) d\tau   +  \nabla V \left( x - \int_s^t \phi^x (\tau) d \tau \right)  \int_s^t v^x (\tau) d \tau \ ds
\nonumber\\
&=&   \int_0^t   \phi^p (s)  \int_s^t  v^x (\tau) d\tau  +  \nabla V \left( x -   \int_s^t \phi^x (\tau) d \tau \right)  \int_s^t v^x (\tau) d \tau \ ds .
\nonumber
\end{eqnarray}
This entails:
\begin{equation}
\label{eq-pf01} 
\phi^p (s) =  - \nabla V \left( x -   \int_s^t \phi^x (\tau) d \tau \right) 
\end{equation}
that is equivalent to
\begin{equation}
\dot{\gamma}^p (t,x,\phi) (s)= - \nabla_x V ( \gamma^x (t,x,\phi) (s)). 
\end{equation}
By a simple computation and (\ref{eq-pf00}) we get: 
\begin{eqnarray}
\nabla_\eta \mathcal{S} &=& x - t \frac{\eta}{m} - \frac{1}{m }  \int_0^t \int_0^s   \phi^p (\tau) d\tau ds
= x - t \frac{\eta}{m} -  \int_0^t  \left(  \phi^x (s) -  \frac{1}{m} \eta \right) ds
\nonumber\\
&=& x - t \frac{\eta}{m}  -    \int_0^t  \phi^x (s) ds + t \frac{\eta}{m} 
=  x -   \int_0^t  \phi^x (s) ds =  \gamma^x (t,x,\phi)(0) = y
\nonumber
\end{eqnarray}
Finally, by (\ref{eq-pf01}), we can complete the verification: 
\begin{eqnarray}
\nabla_x  \mathcal{S}  &=& \eta -  \int_0^t     \nabla V \left( x  -  \int_s^t  \phi^x (\tau)  d\tau \right)   ds
\nonumber\\
&=& \eta +  \int_0^t   \phi^p (s) ds =  \eta +  \gamma^p (t,x,\phi) (t) = p
\nonumber
\end{eqnarray}
\endproof

The following statement is a direct consequence of the above result:
\begin{proposition}
\label{eq-hj-s}
The Hamilton-Jacobi  equation is solved on the stationarity points $\displaystyle 0=  \frac{ D \mathcal{S} }{D\phi}$; more precisely $\mathcal{S}$ is a smooth solution of the problem: 
\begin{equation}
\label{eqHJ1} \left\{
\begin{array}{l}
\displaystyle{ \partial_t \mathcal{S}(t,x,\eta,\phi)  +  \frac{|\nabla_x \mathcal{S}|^2}{2m}(t,x,\eta,\phi)  + V(x) = 0, \quad (t,x) \in \Bbb R^+ \times \Bbb R^n }
\\
\\
\displaystyle{\mathcal{S}(0,x,\eta,\phi) = \langle x , \eta \rangle, \quad   \frac{ D \mathcal{S} }{D\phi} (t,x,\eta,\phi) = 0.}
\end{array}
\right.
\end{equation}
\end{proposition}
\noindent
For the proof we refer to  \cite{C2} (sections 3 and 4). 
\begin{remark}
\label{mapG}
As we have seen in Proposition \ref{prop-sv0}, fix $t \in [0,T]$ and define the map 
\begin{eqnarray}
\label{Gt1}
&&
G_t : \Bbb R^{2n} \times L^2 ([0,T];\Bbb R^{2n}) \rightarrow  L^2 ([0,T];\Bbb R^{2n})
\\
&&
\label{Gt2}
G_t (x,\eta,\phi^x,\phi^p)  := \left( \frac{\eta}{m} + \frac{1}{m} \int_0^s \phi^p (\tau) \ d\tau,  - \nabla V \left( x - \int_s^t \phi^x (\tau) d \tau \right) \right).
\end{eqnarray}
Then, the fixed point equation on $L^2 ([0,T];\Bbb R^{2n})$: 
\begin{equation}
\label{eqfix-01}
\phi  = G_t (x,\eta,\phi) \quad \quad \quad 
\end{equation}
is equivalent to the stationarity equation 
$$
0 = \frac{ D \mathcal{S} }{D\phi}(t,x,\eta,\phi)
$$
On the other hand, the solution of this equation determines the curves 
\begin{eqnarray}
\zeta(t,x,\phi)(s)  := \left(  x  -   \int_s^t  \phi^x (\tau) d\tau  , \eta +
\int_0^s \phi^p (\tau) d\tau  \right) \quad \quad  \phi \equiv (\phi^x,\phi^p)
\end{eqnarray}
solving the Hamilton's equations $\dot{\zeta} = J \nabla H (\zeta)$ with boundary conditions $\zeta^x (t) = x$,  $\zeta^p (0) = \eta$.
\end{remark}

The following result deals with some topological properties for the set of the solutions. Let:
\begin{eqnarray}
&&
\label{lambda}
t_{\alpha,\beta} := \frac{\pi}{2\sqrt{\lambda_\alpha}}(2\beta +1), \;\alpha=1,2,\ldots,n;\quad \beta\in \mathbb{N}
\\
&& 
\label{risonanze}
\lambda(x,\eta) := \sqrt{1+ |x|^2 + |\eta|^2}
 \end{eqnarray}
 where $\lambda_\alpha: \alpha=1,2,\ldots,n$ are the eigenvalues of $L+L^\dag$. Remark that $t_{\alpha,\beta}$ are just the resonant times of the hamiltonian flow generated by  $\ds \mathcal{H}_0 :=  \frac{p^2}{2m} + \langle Lx,x \rangle$.  
\begin{proposition}
\label{bound-gen}
There are  $D(T) < + \infty$, $K_2(T) < + \infty$,  $K_1(t) < +\infty$ such that the solutions of equation (\ref{eqfix-01}) fulfill the estimates:
\begin{eqnarray}
\label{first-ine}
\| \phi \|_{L^2} &>&   K_2 (T) \lambda(x,\eta) \quad \forall t \in ]0,T],  \quad \   |x|^2+|\eta|^2 > D(T)^2;
\\
\label{second-ine}
\| \phi \|_{L^2} &\le&   K_1 (t) \lambda(x,\eta)  \quad \forall t \neq t_{\alpha,\beta}, \quad \forall (x,\eta) \in \Bbb R^{2n}.
\end{eqnarray} 
 Moreover, there is $E(t)<+\infty$ such that  the difference of  any two solutions $\phi,\psi$ of (\ref{eqfix-01})  fulfills the estimate
\begin{eqnarray}
\label{third-ine}
\|  \phi - \psi  \|_{L^2} \le E(t) \quad  \forall t \neq t_{\alpha,\beta}, \quad \forall (x,\eta) \in \Bbb R^{2n}.
\end{eqnarray} 
\end{proposition}
\proof
We begin by remarking that the equation (\ref{eqfix-01})
$$
(\phi^x,\phi^p) = \left( \frac{\eta}{m} + \frac{1}{m} \int_0^s \phi^p (\tau) \ d\tau,  - \nabla V \left( x - \int_s^t \phi^x (\tau) d \tau \right) \right),
$$
can be rewritten as 
\begin{eqnarray}
\label{ex-fixeqn}
\phi - \mathcal{L}(t,\phi) = \Psi_0 (t,x,\eta) + \Psi_1 (t,x,\phi)
\end{eqnarray} 
where
\begin{eqnarray}
\mathcal{L}(t,\phi) &:=&  \left( \frac{1}{m} \int_0^s \phi^p (\tau) \ d\tau,   (L+L^{\dag})    \int_s^t \phi^x (\tau) d \tau  \right)
\nonumber\\
\Psi_0 (t,x,\eta)     &:=&  \left( \frac{\eta}{m} , - (L+L^{\dag})x \right) 
\nonumber\\
\Psi_1 (t,x,\phi)    &:=&  \left( 0 , - \nabla V_0 \left( x - \int_s^t \phi^x (\tau) d \tau \right) \right) 
\end{eqnarray} 
To prove the inequality (\ref{first-ine}), remark that the non-degeneracy of $L+L^{\dag}$ entails the lower and upper bounds:
\begin{eqnarray}
\label{stima-Psi0}
W_0 (T) \lambda(x,\eta) \le \|  \Psi_0 (t,x,\eta) \|_{L^2} = T^{\frac{1}{2}} \left( \frac{|\eta|^2}{m^2} +  |(L+L^{\dag})x|^2  \right)^{\frac{1}{2}}   \le  C_0 (T) \lambda(x,\eta).
\end{eqnarray} 
Here $W_0 (T) := T^{\frac{1}{2}} \mu_M$, $C_0 (T) := T^{\frac{1}{2}};  \mu_m$ and  $\mu_M,\mu_m$ are the maximum and the minimum eigenvalue of the matrix:
\begin{displaymath}
\mathbf{X} = \left( 
\begin{array}{cc}
\frac{1}{m^2} I & 0  \\ 0 & (L+L^{\dag})^2   
\end{array} 
\right) 
\end{displaymath}
respectively. Moreover,
\begin{eqnarray}
\label{stima-Psi1}
\|  \Psi_1 (t,x,\phi)  \|_{L^2} =   \left(  \int_0^T  |\nabla V_0 \left( x - \int_s^t \phi^x (\tau) d \tau \right)|^2   ds \right)^{\frac{1}{2}} \le T^{\frac{1}{2}} \|  \nabla V_0 \|_{C^0} =: C_1 (T)
\end{eqnarray}
Now set $\mathcal{M}(t,\phi) := \phi - \mathcal{L}(t,\phi)$.  Hence the solutions of the equation  (\ref{ex-fixeqn})  fulfill the estimate:
\begin{eqnarray}
\sup_{ t \in [0,T] }  \| \mathcal{M}(t,\cdot) \|_{L^2 \mapsto L^2} \| \phi \|_{L^2} &\ge&  \|  \mathcal{M}(t,\phi)    \|_{L^2} = \|  \Psi_0 (t,x,\eta) + \Psi_1 (t,x,\phi) \|_{L^2} 
\nonumber\\
&\ge& W_0 (T) \lambda(x,\eta) - C_1(T)
\end{eqnarray}
For $|x|^2+|\eta|^2 > D(T)^2$ we have $\ds W_0 (T) \lambda(x,\eta) - C_1(T) >  \frac{W_0 (T)}{2} \lambda(x,\eta)$, and this implies
$$
\| \phi \|_{L^2}  > \left(  \sup_{t\in [0,T]}  \| \mathcal{M}(t,\cdot) \|_{L^2 \mapsto L^2}  \right)^{-1} \frac{W_0 (T)}{2} \lambda(x,\eta)  =: K_2 (T)   \lambda(x,\eta)
$$
Now consider equation (\ref{ex-fixeqn}) in the particular case of $V_0 = 0$ (so that the Hamiltonian is $\ds \mathcal{H}_0 :=  \frac{p^2}{2m} + \langle Lx,x \rangle$). It becomes:
\begin{eqnarray}
\label{griffin}
\phi - \mathcal{L}(t,\phi) = \Psi_0 (t,x,\eta)  \quad \quad  (x,\eta) \in \Bbb R^{n} \times \Bbb R^n
\end{eqnarray}  
The explicit representation of the flow for the harmonic oscillator is $\phi_{\mathcal{H}_0}^s (\zeta^x_0,\zeta^p_0) = e^{s U} (\zeta^x_0,\zeta^p_0)$ where
\begin{displaymath}
\mathbf{U} = \left( 
\begin{array}{cc}
0  & \frac{I}{m}  \\ - (L+L^{\dag}) & 0    
\end{array} 
\right) 
\end{displaymath}
It is easy to prove that outside the resonant times $t_{\alpha,\beta}$ the flow can be globally inverted with respect to the boundary conditions $(x,\eta)$, namely:  $\zeta^x (s) = \zeta^x (t,x,\eta) (s)$ and  $\zeta^p (s) = \zeta^p (t,x,\eta) (s)$. 
By recalling Remark \ref{mapG}, this fact is equivalent to the existence of a unique global smooth solution $\phi^\star_0 (t,x,\eta)$ for equation (\ref{griffin}).  This argument works $\forall (x,\eta) \in \Bbb R^{n} \times \Bbb R^n$. In the particular case $x=\eta=0$ (\ref{griffin}) reduces to:
\begin{eqnarray}
\mathcal{M}(t,\phi) := \phi - \mathcal{L}(t,\phi) = 0 
\end{eqnarray}  
The uniqueness of the solution implies that the linear operator $\mathcal{M}(t,\cdot): L^2 ([0,t];\Bbb R^{2n}) \rightarrow L^2 ([0,t];\Bbb R^{2n})$, $t \neq t_{\alpha,\beta}$, is invertible. Now, we can come back to the general equation (\ref{ex-fixeqn}), written in the equivalent form:
\begin{eqnarray}
\label{park}
\phi =   \mathcal{M}^{-1}(t,   \Psi_0 (t,x,\eta)  +  \Psi_1 (t,x,\phi)  )
\end{eqnarray} 
for all  $t \neq t_{\alpha,\beta}$. By (\ref{stima-Psi0}) and (\ref{stima-Psi1}), we get the inequality
\begin{eqnarray}
\| \phi \| \le   \| \mathcal{M}^{-1}(t,\cdot)\|_{L^2 \mapsto L^2}  (   C_0 (T) \lambda(x,\eta) +  C_1 (T)  )  \le K_1 (t) \lambda(x,\eta)
\end{eqnarray} 
where $K_1 (t):= \| \mathcal{M}^{-1}(t,\cdot)\|_{L^2 \mapsto L^2}  (   C_0 (T) +  C_1 (T)  )$. Finally, we have to prove the bound for the difference of any two solutions $\phi,\psi$ for the equation (\ref{park}). In order to do this, we rewrite it under the form  
$$
\phi - \mathcal{M}^{-1}(t,  \Psi_1 (t,x,\phi)  ) =   \mathcal{M}^{-1}(t,   \Psi_0 (t,x,\eta)  )
$$
As a consequence,
$$
\phi - \psi = \mathcal{M}^{-1}(t,  \Psi_1 (t,x,\phi)  )  -   \mathcal{M}^{-1}(t,  \Psi_1 (t,x,\psi)  ) .  
$$
Recalling (\ref{stima-Psi1}) we have 
$$
\| \phi - \psi \|_{L^2} \le \|  \mathcal{M}^{-1}(t,   )\|_{L^2 \mapsto L^2}   \|    \Psi_1 (t,x,\phi)  - \Psi_1 (t,x,\psi)  \|_{L^2} \le \|  \mathcal{M}^{-1}(t,   )\|_{L^2 \mapsto L^2}  2 C_1 (T) =: E(t)
$$
and this concludes the proof.
\endproof

\subsection{Generating function with finitely many parameters}
In this section we describe how the global parametrization of the graph $\Lambda_t$ of the Hamiltonian flow  can be actually obtained through  a  generating function with finitely many parameters. To this end we use the reduction of the Hamilton-Helmholtz functional due to Amann, Conley and Zehnder  (see \cite{A-Z}, \cite{C-Z1}, \cite{C2}, \cite{C4}). In this way we find, for the graph $\Lambda_t$ of the Hamiltonian flow,  a global parametrization of type: 
$$
\Lambda_t = \left\{ (y,\eta;x,p) \in T^\star \Bbb R^n \times T^\star \Bbb R^n \ | \  p = \nabla_x S, \quad y =  \nabla_\eta S, \quad   0 =  \nabla_\theta S ( t,x,\eta,\theta)  \right\}  
$$
The essence of the Amann-Conley-Zehnder reduction is  the existence of an underlying  finite dimensional structure for the equation investigated in the previous section:
\begin{equation}
\label{eqfix-02}
(\phi^x,\phi^p)  = G_t (x,\eta,\phi^x,\phi^p) \quad \quad \quad (\phi^x,\phi^p) \in L^2 ([0,T];\Bbb R^{2n}).
\end{equation}
We can indeed consider the two orthogonal projectors  
$$
\mathbb{P}_M \phi(s) = \sum_{|r| \le M} \phi^{(r)} e_r (s) \quad \quad \mathbb{Q}_M \phi(s) = \sum_{|r| > M} \phi^{(r)} e_r (s)
$$
generated by any orthonormal basis of $L^2 ([0,T];\Bbb R^{2n})$; for instance  $e_r (s) := \frac{1}{\sqrt{T}} e^{ \frac{2\pi}{T} i r s}: r\in {\mathbb Z}$.  Then, let us introduce the decomposition:
\begin{eqnarray}
\label{fix0}
( f^x(\theta), f^p(\theta) )  &=& \mathbb{Q}_M G_t \left(x,\eta,\theta^x+ f^x (\theta),\theta^p + f^p (\theta)  \right) 
\\
(\theta^x,\theta^p)  &=& \mathbb{P}_M G_t  \left(x,\eta,\theta^x+ f^x (\theta),\theta^p + f^p (\theta)  \right)
\label{fix1}
\end{eqnarray}
and prove the following
\begin{lemma}
\label{corr-sol}
For $M\in \Bbb N$ large enough the functional equation (\ref{fix0}) admits  a unique  solution $f(\theta):  \mathbb{P}_M L^2 \rightarrow  \mathbb{Q}_M L^2$.   The solutions of (\ref{eqfix-02}) can then be written in the form  
$$
( \phi^x , \phi^p ) = (\theta^x+ f^x (\theta), \theta^p+ f^p (\theta) )
$$
where $\theta \in \mathbb{P}_M L^2 \simeq \Bbb R^k$ are finite dimensional parameters solving the fixed point equation (\ref{fix1}) on $\Bbb R^k$, $k=2n(2M+1)$.
\end{lemma}
\proof
Let us first verify that, if $M \in \Bbb N$ is large enough, the equation (\ref{fix0})  realizes in fact a contraction on $C^0(L^2 , L^2)$. Hence it admits a unique solution $f (t,x,\theta)=(f^x (t,x,\theta),f^p (t,x,\theta))$. By (\ref{Gt1}), (\ref{Gt2}) the  two equations read:
$$
(f^x  (t,x,\theta)(s) , f^p (t,x,\theta)(s))  =  \mathbb{Q}_M \left( \frac{1}{m} \int_0^s  f^p (\theta)(\tau) \ d\tau,  -    \nabla V \left( x -   \int_s^t \theta^x (\tau) + f^x (\theta)(\tau)  d \tau \right) \right)
$$
$$
(\theta^x (s),\theta^p(s))  =  \mathbb{P}_M \left( \frac{\eta}{m}  +\frac{1}{m  } \int_0^s \theta^p (\tau) + f^p (\theta)(\tau) \ d\tau,  -    \nabla V \left( x  -   \int_s^t \theta^x (\tau) + f^x (\theta)(\tau)  d \tau \right) \right)
$$
\vskip 5pt\noindent
 It is  proved in \cite{C2} (Lemma 6) that the contraction property holds if: 
 \begin{equation}
T^2  \sup_{(x,p) \in T^\star \Bbb R^n} |\nabla^2 {\mathcal H} (x,p)| \  \frac{1+ \sqrt{2M}}{2 \pi M}  <  1
\end{equation}
By Definition 2.4 it follows that $\sup_{(x,p) \in T^\star \Bbb R^n} |\nabla^2 {\mathcal H} (x,p)| < + \infty$ and  consequently given $0<T<\infty$ we get the contraction property for the first equation choosing $M(T)$ large enough. In general, the second equation have many solutions depending on the values of $(t,x)$.
\endproof

\indent 
In this finite dimensional setting, we can consider the following set of curves, with $t \in [0,T]$:
\begin{equation}
\label{fin-curv} \left\{
\begin{array}{l}
\displaystyle{\gamma^x (t,x,\theta + f(\theta) )(s) = x  -   \int_s^t  \phi^x (t,x,\theta) (\tau) \ d\tau ,
\quad \quad \phi^x (t,x,\theta) = \theta^x  +  f^x  (t,x,\theta)}, 
\\
\\
\displaystyle{ \gamma^p (t,x,\theta + f(\theta) )(s) =   \int_0^s  \phi^p (t,x,\theta) (\tau) \ d\tau ,
\quad \quad \quad \quad   \phi^p (t,x,\theta) = \theta^p  +  f^p  (t,x,\theta)}
\end{array}
\right.
\end{equation}
We note that  this is a finite reduction of (\ref{set-gen}), but still  contains all curves solving Hamilton's equations with boundary data $\gamma^x (t) = x$ and $\gamma^p (0) = 0$ because of $\phi$ are solving equation (\ref{eqfix-02}). Moreover, by a Sobolev's immersion theorem,  $\Gamma \subset H^1([0,T]; T^{\star} \Bbb R^{n}) \subset C^0 ([0,T]; T^{\star} \Bbb R^{n})$ and this entails their continuity.

We can now proceed to define the main object of this section:
\begin{definition}
\label{def-gen-f}
The finitely-many parameters generating function of $\Lambda_t$ is defined as:
\begin{eqnarray}
\label{def-gen-34}
S(t,x,\eta,\theta) &:=&   \langle x, \eta \rangle +   \int_0^t  \gamma^p (s)  \dot{\gamma}^x (s)  -  H(\gamma^x (s), \eta + \gamma^p(s)) \ ds \Big|_{\gamma (\cdot) = \gamma(t,x,\theta+ f(t,x,\theta))(\cdot)}
\nonumber\\
&=&  \mathcal{S} (t,x,\eta,\theta + f(t,x,\theta)) .
\end{eqnarray}
Here $\mathcal{S}$ is the infinite dimensional generating function of Definition 2.3. 
\end{definition}
\noindent
{\bf Remark}
The above generating function is fully parametrized by $\theta + f(t,x,\theta)$, $\theta\in {\Bbb R}^k$, and not by an arbitrary $\phi \in L^2$. This is the core of the finite reduction. 

Now we provide a more detailed study about the analytical properties of $f$. 
\begin{lemma}
\label{stime-f11}
Consider the pair of functions $(f^x,f^p)$. Then:
\par\noindent
(1) $(f^x,f^p)$ fulfill the following equations
\begin{eqnarray}
f^x (s)  &=& \frac{1}{m} \mathbb{Q}_M \int_0^s \mathbb{Q}_M \int_\tau^t (L + L^{\dag})  f^x (r)  d r d\tau +
\Phi^x (t,x,\theta,f^x)(s)
\nonumber\\
f^p (s)   &=&    \mathbb{Q}_M  \int_s^t  (L + L^{\dag})   f^x (\tau)  d\tau + \Phi^p (t,x,\theta,f^x)(s)
\nonumber\\
\label{Phi}
\Phi^x (t,x,\theta,f^x)(s) &:=& - \frac{1}{m} \mathbb{Q}_M \  \int_0^s        \mathbb{Q}_M    \nabla V_0 \left( x -   \int_\tau^t \theta^x (r) + f^x (r) d r \right) d\tau .
\nonumber\\
\Phi^p (t,x,\theta,f^x)(s) &:=& -    \mathbb{Q}_M    \nabla V_0 \left( x -   \int_s^t \theta^x (r) + f^x (r) d r \right) 
\end{eqnarray}
(2) Under the condition 
\begin{equation}
d := \frac{T^2}{m}  \|   \mathbb{Q}_M  \|^2 \|  L + L^{\dag} \|  < 1,
\end{equation}
 they fulfill the estimates:
\begin{eqnarray}
\|  f^x (t,x,\theta) (\cdot) \|_{L^2} &\le&  \left( 1-  d  \right)^{-1}    \frac{ T^\frac{3}{2} }{m}  \|  \mathbb{Q}_M   \|^2  \| \nabla V_0 \|_{C^0} 
\nonumber\\
\|  f^p (t,x,\theta) (\cdot) \|_{L^2} &\le& T \| L \|   \|   \mathbb{Q}_M  \|  \|  f^x (t,x,\theta) (\cdot) \|_{L^2} +  \frac{T^{\frac{1}{2}} }{m}  \|   \mathbb{Q}_M  \|  \| \nabla V_0 \|_{C^0}
\end{eqnarray}
(3) If in addition  
\begin{equation}
\label{bound-M}
\frac{T^2}{m}  \|   \mathbb{Q}_M  \|^2     \sup_{|i|+|j| \ge 2}\sup_{x,p}    |\partial^i_x \partial^j_p {\mathcal H} (x,p) |^2 < 1
\end{equation}
 then there exist $C_{\alpha \sigma} (T)>0$ such that:
\begin{eqnarray}
\label{first-partialb}
\|  \partial^\alpha_x \partial^\sigma_\theta   f (t,x,\theta) (\cdot) \|_{L^2} \le C_{\alpha \sigma} (T)
\end{eqnarray}
\end{lemma}
\proof
By direct computation, the first functional equation reads: 
\begin{eqnarray*}
f^x (s)  &=&  \frac{1}{m} \mathbb{Q}_M \   \int_0^s  f^p (\theta)(\tau) \ d\tau
\\
&=&  \frac{1}{m} \mathbb{Q}_M \  \int_0^s   -   \mathbb{Q}_M    \nabla V \left( x -   \int_\tau^t \theta^x (r) + f^x (r)  dr \right)  \ d\tau
\\
&=& - \frac{1}{m} \mathbb{Q}_M \  \int_0^s     \mathbb{Q}_M    (L + L^{\dag}) \left( x -   \int_\tau^t \theta^x (r) +  f^x (r) d r \right)  d\tau   
\\
&-&  \frac{1}{m} \mathbb{Q}_M \int_0^s     \mathbb{Q}_M  \nabla V_0 \left(  x  -  \int_\tau^t \theta^x (r) + f^x (r) dr \right)  d\tau
\\
&=&   \frac{1}{m} \mathbb{Q}_M \int_0^s \mathbb{Q}_M \int_\tau^t (L + L^{\dag})  f^x (r)  d r d\tau +
\Phi^x  (t,x,\theta,f^x)(s) 
\end{eqnarray*}
where the last equality follows by (\ref{Phi}). Analogous computation for $f^p(s)$:
$$
f^p (t,x,\theta)(s)   =    \mathbb{Q}_M  \int_s^t  (L + L^{\dag})   f^x (\tau)  d\tau + \Phi^p (t,x,\theta,f^x)(s) 
$$
This proves Assertion (1).
\par\noindent
To see Assertion (2), remark that (\ref{Phi}) also entails:
\begin{eqnarray}
\| \Phi^x (t,x,\theta,f^x)(\cdot) \|_{L^2} \le   \frac{T^{\frac{3}{2}} }{m}  \|   \mathbb{Q}_M  \|^2  \| \nabla V_0 \|_{C^0} 
\end{eqnarray}
whence we obtain:
\begin{eqnarray*}
\|  f^x (t,x,\theta) (\cdot) \|_{L^2} &\le&  \frac{T^2}{m}  \|   \mathbb{Q}_M  \|^2 \|   L + L^{\dag} \|   \|  f^x (t,x,\theta) (\cdot) \|_{L^2}   
+  \| \Phi (t,x,\eta,\theta)(\cdot) \|_{L^2}
\end{eqnarray*}
If we choose  $M$ large enough, then $d := \frac{T^2}{m}  \|   \mathbb{Q}_M  \|^2 \|   L + L^{\dag} \| < 1$ and hence we get
\begin{eqnarray*}
\|  f^x (t,x,\theta) (\cdot) \|_{L^2} &\le& \left( 1- d  \right)^{-1}    \frac{T^{\frac{3}{2}} }{m}  \|   \mathbb{Q}_M  \|^2  \| \nabla V_0 \|_{C^0} 
\nonumber
\end{eqnarray*}
In the same way we have the estimate:
\begin{eqnarray*}
\|  f^p (t,x,\theta) (\cdot) \|_{L^2}  &\le& T^{\frac{1}{2}}  \|  L + L^{\dag}  \|   \|   \mathbb{Q}_M  \| \|  f^x (t,x,\theta) (\cdot) \|_{L^2}   +     \| \Phi^p (t,x,\eta,\theta)(\cdot) \|_{L^2}
\nonumber\\
&\le& T^{\frac{1}{2}}  \|  L + L^{\dag}  \|   \|   \mathbb{Q}_M  \| \|  f^x (t,x,\theta) (\cdot) \|_{L^2}   +    T^{\frac{1}{2}}    \|   \mathbb{Q}_M  \|   \| \nabla V_0 \|_{C^0} 
\end{eqnarray*}
This proves Assertion (2).
\par\noindent
The equation for the first order partial derivatives reads:
\vskip 4pt\noindent
\begin{eqnarray*}
\lefteqn{
\frac{\partial f^{x,\alpha}}{\partial x_i}  (t,x,\theta) =  \frac{\mathbb{Q}_M}{m}  \int_0^s \mathbb{Q}_M \int_\tau^t (L + L^{\dag})  \frac{\partial f^{x,\alpha}}{\partial x_i}  (t,x,\theta) (r)  d r d\tau }
\nonumber\\
&+& \frac{\mathbb{Q}_M  }{m}  \int_0^s  \mathbb{Q}_M   \frac{ \partial^2 V_0}{\partial x_\alpha \partial x_\beta} \left( x -   \int_\tau^t \theta^x (r) + f^x (t,x,\theta) (r) dr \right) \left( \delta_{\beta i} +  \int_\tau^t \frac{\partial f^{x,\beta}}{\partial x_i}  (t,x,\theta)(r) dr \right)  d\tau
\end{eqnarray*}
If $M$ is large enough, then $d^{\prime}:= \frac{T^2}{m}  \|   \mathbb{Q}_M  \|^2 \|   L + L^{\dag} \|  +  \frac{T^2}{m}  \|   \mathbb{Q}_M  \|^2  \| \nabla^2 V_0 \|_{C^0}<1$ and we get:
$$
\left\| \frac{\partial f^x}{\partial x_i}  (t,x,\theta) (\cdot) \right\|_{L^2} <  (1 - d^{\prime})^{-1}  \frac{T^{\frac{3}{2}}}{m}    \|  \mathbb{Q}_M  \|^2  \| \nabla^2 V_0 \|_{C^0}
$$
The equation for the second order partial derivatives reads:
\begin{eqnarray*}
\lefteqn{
\frac{\partial^2 f^{x,\alpha}}{\partial x_i \partial x_j} (t,x,\theta) =  \frac{\mathbb{Q}_M }{m} \int_0^s \mathbb{Q}_M \int_\tau^t (L + L^{\dag})  \frac{\partial^2 f^{x,\alpha}}{\partial x_i \partial x_j}  (t,x,\theta) (r)  dr d\tau }
\nonumber\\
&+& \frac{\mathbb{Q}_M }{m}   \int_0^s  \mathbb{Q}_M     \frac{ \partial^2 V_0}{\partial x_\alpha \partial x_\beta}  \left( x -   \int_\tau^t \theta^x (r) + f^x (t,x,\theta) (r) dr \right) \left(  \int_\tau^t \frac{\partial^2 f^{x,\beta}}{\partial x_i \partial x_j}  (t,x,\theta)(r) dr \right)d\tau 
\nonumber\\
&+& \frac{\mathbb{Q}_M }{m}   \int_0^s  \mathbb{Q}_M    F_{\alpha \beta k}(t,x,\theta)(\tau)  \left( \delta_{kj} +  \int_\tau^t \frac{\partial f^{x,k}}{\partial x_j} (t,x,\theta)(r) dr \right)  \left( \delta_{\beta i} +  \int_\tau^t \frac{\partial f^{x,\beta}}{\partial x_i}  (t,x,\theta)(r) dr \right)d\tau 
\nonumber\\
\end{eqnarray*}
where
$$
F_{\alpha \beta k}(t,x,\theta)(\tau) :=   \frac{ \partial^3 V_0}{\partial x_\alpha \partial x_\beta  \partial x_k}  \left( x -   \int_\tau^t \theta^x (r) + f^x (t,x,\theta) (r) dr \right)
$$
As before,  if we require $d^{\prime \prime}:=  \frac{T^2}{m}  \|   \mathbb{Q}_M  \|^2 (  \|   L + L^{\dag} \|  +  \| \nabla^2 V_0 \|_{C^0} + 
 \|  F \|_{C^0}  ) < 1$ then
$$
\left\| \frac{\partial^2 f^x}{\partial x_i \partial x_j}   (t,x,\theta) (\cdot) \right\|_{L^2}  <  (1-d^{\prime \prime})^{-1}   \frac{ 1 }{m} \|   \mathbb{Q}_M  \|^2 \|  F \|_{C^0}  \left( T^{\frac{3}{2}}    + T^4  \|  \nabla f^x \|^2_{L^2} + 2 T^2  \|  \nabla f^x \|_{L^2} \right)
$$
For the higher order derivatives in $x$ and also for $\theta$-partial derivatives we can proceed in the same way, with the general condition
\begin{eqnarray}
\frac{T^2}{m}  \|   \mathbb{Q}_M  \|^2   \sup_{|i|+|j| \ge 2 }\sup_{x,p}    |\partial^i_x \partial^j_p {\mathcal H} (x,p) |^2 < 1 
\end{eqnarray}
in order to conclude the existence of $C_{\alpha \sigma} (T)>0$ such that
\begin{eqnarray}
\|  \partial^\alpha_x \partial^\sigma_\theta   f (t,x,\theta) (\cdot) \|_{L^2} \le C_{\alpha \sigma} (T).
\end{eqnarray}
This proves Assertion (3) and thus concludes the proof of the Lemma.
\endproof

\begin{theorem}
\label{Rep-S}
The generating function (\ref{def-gen-34}) admits the following representation:
\begin{eqnarray}
\label{rep-gen99}
S &=&  \langle x , \eta \rangle   - \frac{t}{2m} \eta^2 -  t \langle L x, x \rangle   +   \langle Q(t)  \theta  ,   \theta   \rangle 
+ \langle v(t,x,\eta) , \theta + f(t,x,\theta)  \rangle + \langle \nu(t,x,\theta) , \theta  \rangle 
\nonumber\\
&+&  g (t,x,\theta).
\end{eqnarray}
Here:  $\theta \in \Bbb R^k$; $t\mapsto Q(t)\in GL(n)$, $Q(0)=0$; more overthere are $C_{\alpha \beta \sigma}(T)>0$ such that 
$$
|\partial_x^\alpha \partial_\eta^\beta \partial_\theta^\sigma g |+|\partial_x^\alpha \partial_\eta^\beta \partial_\theta^\sigma \nu |+|\partial_x^\alpha \partial_\eta^\beta \partial_\theta^\sigma f | \le C_{\alpha \beta \sigma}(T). 
$$
The function $(t,x,\eta)\mapsto v(t,x,\eta)$ is linear in $x,\eta$, and finally:
\begin{eqnarray}
k >   C  T^4   \sup_{|\alpha| + |\beta| \ge 2} \,\sup_{x,p} \,   |\partial^\alpha_x \partial^\beta_p {\mathcal H} (x,p) |^2.
\end{eqnarray} 
\end{theorem}
\proof
By the explicit upper bound $\displaystyle \| \mathbb{Q}_M \| \le \frac{T}{2\pi} \sqrt{\frac{2}{M}}$  of (\ref{bound-M}) we have
$$
k >   C  T^4   \sup_{|\alpha| + |\beta| \ge 2} \,\sup_{x,p} \,   |\partial^\alpha_x \partial^\beta_p {\mathcal H} (x,p) |^2.
$$ 
We recall the structure of the infinite dimensional generating function:
$$
\mathcal{S} (t,x,\eta,\phi) = \langle x , \eta \rangle   - \frac{t}{2m} \eta^2  - t \langle L x, x \rangle   +  \langle R(t) \phi , \phi  \rangle +  \langle v(t,x,\eta) , \phi \rangle + \sigma (t,x,\phi) 
$$
As  a consequence: 
\begin{eqnarray}
S(t,x,\eta,\theta) &:=&  \mathcal{S} (t,x,\eta,\theta + f(t,x,\theta)) 
\nonumber\\
&=& \langle x , \eta \rangle   - \frac{t}{2m} \eta^2 - t \langle L x, x \rangle   +  \langle R(t) ( \theta + f(t,x,\theta) ) , \theta + f(t,x,\theta) \rangle 
\nonumber\\
&+&  \langle v(t,x,\eta) , \theta + f(t,x,\theta) \rangle  + \sigma (t,x,\theta + f(t,x,\theta)) . 
\nonumber
\end{eqnarray}
We can thus make the identifications:
\begin{eqnarray}
\langle Q(t)  \theta  ,   \theta   \rangle &:=& \langle R(t) \theta , \theta  \rangle
\nonumber\\
\langle \nu(t,x,\theta) , \theta \rangle &:=&  \langle 2 R(t)  f(t,x,\theta) , \theta \rangle
\nonumber\\
g (t,x,\theta) &:=&  \sigma (t,x,\theta + f(t,x,\theta))  +   \langle R(t)  f(t,x,\theta)  ,  f(t,x,\theta) \rangle
\end{eqnarray}
Now it is easy to see that
$$
| \nu(t,x,\theta) | \le 2 \| R(t) \| \| f(t,x,\theta)(\cdot) \|_{L^2} \le C(T)
$$
and this entails boundedness with respect to the $\theta$-variables. Moreover the same property holds true for the other term. We have indeed:
$$
|g (t,x,\theta)| \le  \| \sigma (t,x,\phi)(\cdot)\|_{C^0}  + \|  R(t) \| \| f(t,x,\theta)(\cdot) \|_{L^2}^2 \le C^\prime (T)
$$
By using the  above results  we then get the existence  of $C^\prime (T)$ such that $|g (t,x,\theta) \le C^\prime (T)$. The estimates for  the partial derivatives follow in the same way.
\endproof

\begin{theorem}
\label{prop-sv10}
The graph of the Hamiltonian flow  
$$
\Lambda_t := \left\{  (y,\eta;x,p) \in T^\star \Bbb R^n \times T^\star \Bbb R^n \ | \ (x,p) = \phi_\H^t (y,\eta)     \right\}
$$ 
admits a  global generating function with finitely many parameters:
$$
\Lambda_t = \left\{ (y,\eta;x,p) \in T^\star \Bbb R^n \times T^\star \Bbb R^n \ | \  p = \nabla_x S, \quad y =  \nabla_\eta S, \quad   0 =  \nabla_\theta S(t,x,\eta,\theta) \right\}
$$
\end{theorem}
\proof
By Proposition (\ref{prop-sv0}) we can write:
$$
\Lambda_t = \left\{ (y,\eta;x,p) \in T^\star \Bbb R^n \times T^\star \Bbb R^n \ | \  p = \nabla_x \mathcal{S}, \quad y =  \nabla_\eta \mathcal{S}, \quad   0=  \frac{ D \mathcal{S} }{D\phi} \right\}
$$
where $\mathcal{S}$ is the infinite-dimensional generating function of Definition 2.3. Now we remark that the finite-dimensional stationarity condition:
$$
0 =  \nabla_\theta S(t,x,\eta,\theta^\star)
$$
is equivalent to the variational equation expressing the  stationarity:
$$
0 =  \frac{D\mathcal{S}}{D\phi} (t,x,\eta,\phi^\star) 
$$
Indeed, by Lemma \ref{corr-sol} and $\cite{C2}$ (see Lemma 7), there is a bijective correspondence between the solutions of the two equations,   $\phi^\star = \theta^\star + f (t,x,\theta^\star)$.
Moreover it is easy to prove that
$$
\nabla_x S |_{(t,x,\eta,\theta^\star)} =  \nabla_x \mathcal{S} |_{(t,x,\eta,\phi^\star)}, \quad \quad  \nabla_\eta S |_{(t,x,\eta,\theta^\star)} =  \nabla_\eta \mathcal{S} |_{(t,x,\eta,\phi^\star)}
$$
This is true because of the definition $S(t,x,\eta,\theta) := \mathcal{S} (t,x,\eta,\theta+f(t,x,\theta))$, and the computation
$$
\nabla_x S (t,x,\eta,\theta) =  \nabla_x \mathcal{S} (t,x,\eta,\phi)|_{\phi=\theta+f(t,x,\theta)} +  \frac{D\mathcal{S}}{D\phi} (t,x,\eta,\phi)|_{\phi=\theta+f(t,x,\theta)} [ \nabla_x f(t,x,\theta) ].
$$ 
Evaluating  both sides on the solutions $\theta^\star$ we get the relation. The same argument  applies to  $\nabla_\eta S$,  and this concludes the proof. 
\endproof

\begin{theorem}
\label{eq-hj-r}
The Hamilton-Jacobi  equation is solved by the smooth function $S(t,x,\eta,\theta) $ on the stationary points $\Sigma_S =\{ (x,\eta,\theta) \in R^{2n+k} \ | \  \nabla_\theta S (t,x,\eta,\theta) = 0 \}$. More precisely:
\begin{equation}
\label{eqHJ1} \left\{
\begin{array}{l}
\displaystyle{ \partial_t S(t,x,\eta,\theta) +  \frac{|\nabla_x S|^2}{2m}(t,x,\eta,\theta)  + V(x) = 0,  }
\\
\\
S(0,x,\eta,\theta) = \langle x,\eta\rangle; \quad  (x,\eta,\theta) \in \Sigma_S.
\end{array}
\right.
\end{equation}
\end{theorem}
\proof
\vskip 4pt\noindent
We recall that, by Proposition \ref{eq-hj-s}, the Hamilton-Jacobi  equation is solved by $\mathcal{S}(t,x,\eta,\phi) $ on the infinite dimensional stationary points defined by $\ds  \frac{ D \mathcal{S} }{D\phi} (t,x,\eta,\phi^\star) = 0.$
\begin{equation}
\label{eq-geo11}
\left\{
\begin{array}{l}
\displaystyle{ \partial_t \mathcal{S}(t,x,\eta,\phi)  +  \frac{|\nabla_x \mathcal{S}|^2}{2m}(t,x,\eta,\phi)  + V(x) = 0, \quad (t,x) \in \Bbb R \times \Bbb R^n }
\\
\\
\mathcal{S}(0,x,\eta,\phi) = \langle x . \eta \rangle, \quad{\ds  \frac{ D  \mathcal{S} }{D\phi} (t,x,\eta,\phi) = 0.}
\end{array}
\right.
\end{equation}
On the other hand, we have
\begin{equation}
\label{rel-S11}
\nabla_x S |_{(t,x,\eta,\theta^\star)} =  \nabla_x \mathcal{S} |_{(t,x,\eta,\phi^\star)}, \quad \quad  \partial_t S |_{(t,x,\eta,\theta^\star)} =  \partial_t \mathcal{S} |_{(t,x,\eta,\phi^\star)}
\end{equation}
Indeed the first equality is proved in the previous theorem; whereas for the second one we observe:
$$
\partial_t S (t,x,\eta,\theta) =  \partial_t \mathcal{S} (t,x,\eta,\phi)|_{\phi=\theta+f(t,x,\theta)} +  \frac{D\mathcal{S}}{D\phi} (t,x,\eta,\phi)|_{\phi=\theta+f(t,x,\theta)} [ \partial_t f(t,x,\theta) ].
$$ 
Since  $\ds \frac{ D \mathcal{S} }{D\phi} (t,x,\eta,\phi^\star) = 0$ the second equality in (\ref{rel-S11}) is proved.   (\ref{eq-geo11}) and   (\ref{rel-S11}) then yield  the assertion.
\endproof

\begin{theorem}
\label{theo-S-eq}
Let  $S$ and $\Sigma_S$ be as in Theorem \ref{eq-hj-r}. Then there exists $\Theta_N \in C^\infty_b ( [0,T]  \times \Bbb R^{2n+k} ; \Bbb R )$ with $\Theta_N|_{\Sigma_S}=0$, such that   an equivalent gene\-rating function $S_N$ is given by the solution of the
problem
\begin{equation}
\label{eq-HJ-geom}
 \left\{
\begin{array}{l} {\ds 
\frac{|\nabla_x  S_N |^2}{2m} (t,x,\eta,\theta)  +  V(x)  +  \partial_t  S_N  (t,x,\eta,\theta) = \Theta_N, }
\\
\\
S_N (0,x,\eta,\theta) = \langle x , \eta \rangle. 
\end{array}
\right.
\end{equation}
Moreover, defining:
\begin{eqnarray}
\label{reg-div}
\Pi (\Theta_N ) :=   {\rm div}_\theta \left( \Theta_N \frac{ \nabla_\theta S_N }{|\nabla_\theta S_N |^2} \right) 
\end{eqnarray}
$S_N$ enjoys the property:
\begin{eqnarray}
\label{reg-div2}
\Pi^{ j } (\Theta_N )  \in C^\infty_b ( [0,T]  \times \Bbb R^{2n+k} ; \Bbb R ) \quad  \quad \forall 1 \le j \le N, \quad N=1,2,\ldots
\end{eqnarray}
 \end{theorem}
\proof 
We remember that $\Sigma_S \subset \Bbb R^{2n+k}$ is a submanifold of dimension $2n$ thanks to the nondegeneracy condition
$$
rk (\nabla^2_{x\theta} S , \nabla^2_{\eta \theta} S , \nabla^2_{\theta \theta} S )|_{\Sigma_S} = max = k = 2n(2N+1)
$$
for some $N\ge1$. We define $z:=(x,\eta,\theta) \in \Bbb R^{2n+k}$ and for any point $\bar{z} \in \Sigma_S$. Define furthermore  $\widetilde{S} $ (not necessarily uniquely) through the conditions:
\begin{eqnarray}
\label{first-D}
 \partial_t  \widetilde{S}  (t,z)  &=&  \partial_t  S (t,\bar{z})  + L(t,z) ,
\\
\label{sec-DD}
\nabla_z  \widetilde{S}    (t,z)  &=&   \nabla_z  S (t,\bar{z}) + F(t,z), \\
\label{third-D}
\Delta_x \widetilde{S}  (t,z)  &=&   \Delta_x  S  (t,\bar{z}) +  G(t,z),
\end{eqnarray}
where $L = (L^x,L^\eta,L^\theta) \in C^\infty_b ([0,T] \times \Bbb R^{2n+k}; \Bbb R)$ and $L(t,\bar{z}) = 0$, the perturbation of the gradient in (\ref{sec-DD}) is $F = (F^x,F^\eta, F^\theta)  \in C^\infty_b ([0,T] \times \Bbb R^{2n+k}; \Bbb R^{2n+k})$ with $F(t,\bar{z}) = 0$ while in (\ref{third-D})  we require $G  \in C^\infty_b ([0,T] \times \Bbb R^{2n+k}; \Bbb R)$ and $G(t,\bar{z}) = 0$.  In addition, we require that $F^\theta (t,z) \neq 0$ for $z \notin \Sigma_S$. Hence the new stationarity equation:
\begin{eqnarray}
\nabla_\theta \widetilde{S} (t,z)  =  F^\theta (t,z)   
\end{eqnarray} 
implies  $\Sigma_{\widetilde{S}} = \Sigma_S$. In order to verify (\ref{reg-div}) we require  a suitable asymptotic behaviour of $L,F,G$ around $\Sigma_S$. Indeed,
$$ 
\partial_t  \widetilde{S}  (t,z)  =  \partial_t  S (t,\bar{z})  + L(t,z), \quad  \quad  \nabla_x  \widetilde{S} (t,z) = \nabla_x S (t,\bar{z}) + F^{x}(t,z).      
$$
So, by easy computations and by (\ref{eqHJ1}), we have
\begin{eqnarray}
\Theta (t,z) &=& \frac{|\nabla_x  \widetilde{S} |^2}{2m} (t,z)  +  V(x)  +  \partial_t  \widetilde{S}  (t,z)  
\nonumber\\
&=&  \frac{ 1}{2m} |\nabla_x S (t,\bar{z}) + F^{x}(t,z) |^2  +  V(x)  +   \partial_t  S (t,\bar{z})  + L(t,z)
\nonumber\\
&=&  \frac{1}{2m} |\nabla_x S (t,\bar{z}) |^2  +  V(x)  +   \partial_t  S (t,\bar{z})  +  \frac{1}{m} \nabla_x S (t,\bar{z}) F^{x}(t,z)  + \frac{ 1}{2m} | F^{x}(t,z) |^2  + L(t,z)
\nonumber\\
&=&  \frac{1}{m} \nabla_x S (t,\bar{z}) F^{x}(t,z)  + \frac{ 1}{2m} | F^{x}(t,z) |^2  + L(t,z)
\end{eqnarray} 
Now we can always require that the vanishing asymptotic behaviour of $F^x,F^\theta,L$ around $\Sigma_S$ are such that it holds:
$$
\Pi (\Theta) := {\rm div} \left( \Theta \frac{ \nabla_\theta \widetilde{S}}{|\nabla_\theta \widetilde{S}|^2} \right) \in C^\infty_b ( [0,T]  \times \Bbb R^{2n+k} ; \Bbb R )
$$
By the same arguments as above, we can look for $S_N$ such that $\Sigma_{S_N}  =  \Sigma_S$  and
\begin{eqnarray}
\label{sec-D}
\partial_t  S_N  (t,z)  &=&  \partial_t  S (t,\bar{z})  +  L_N (t,z) ,
\\
\label{sec-D}
\nabla_z  S_N   (t,z)  &=&   \nabla_z  S (t,\bar{z}) + F_N (t,z), \\
\label{third-D}
\Delta_x S_N  (t,z)  &=&   \Delta_x  S  (t,\bar{z}) +  G_N (t,z),
\end{eqnarray}
where  $F^x_N , F^\theta_N$  and $L_N$ are chosen in such a way that:  
$$
\Pi^{ j } (\Theta_N)  \in C^\infty_b ( [0,T]  \times \Bbb R^{2n+k} ; \Bbb R ) \quad  \quad \forall 1 \le j \le N
$$  
\endproof


Let us examine the topology of the finite-dimensional critical points set.
\begin{theorem}
\label{compact}
Let  $S$ be as in Definition \ref{def-gen-f}, and $\lambda(x,\eta)$, $t_{\alpha,\beta}$ as  in (\ref{lambda}) and (\ref{risonanze}), respectively.   Consider  $(t,x,\eta) \in ]0,T] \times \Bbb R^n \times \Bbb R^n$.   Then:
\par\noindent
(1)
 All solutions $\theta \in \mathbb{P}_M L^2([0,T];\Bbb R^{2n}) \simeq \Bbb R^k$ of the stationarity equation:
$$
0 = \nabla_\theta S(t,x,\eta,\theta)
$$
 fulfill  the estimates 
\begin{eqnarray}
\label{dis-theta1}
\| \theta \|_{L^2} &>&   \tilde{K}_2 (T) \lambda(x,\eta) \quad \forall t \in ]0,T],  \quad \   |x|^2+|\eta|^2 > D(T)^2;
\\
\| \theta \|_{L^2} &\le&   \tilde{K}_1 (t) \lambda(x,\eta)  \quad \forall t \neq t_{\alpha,\beta}, \quad \forall (x,\eta) \in \Bbb R^{2n}.
\label{dis-theta2}
\end{eqnarray} 
where $D(T) < + \infty$, $\tilde{K}_2(T) < + \infty$  while $\tilde{K}_1(t) < +\infty$ is a constant  defined for  $t\neq t_{\alpha,\beta} $ .
\par\noindent
(2) The difference of  any two solutions $\theta,\omega$ fulfills the inequality
\begin{eqnarray}
\label{dis-theta3}
\|  \theta - \omega  \|_{L^2} \le \tilde{E}(t) \quad  \forall t \neq t_{\alpha,\beta}, \quad \forall (x,\eta) \in \Bbb R^{2n}.
\end{eqnarray} 
\end{theorem}
\proof
By Proposition  \ref{bound-gen} and  Lemma  \ref{corr-sol} all solutions $\phi \in L^2$of the variational equation
$$
0 = \frac{D\mathcal{S}}{D\phi}(t,x,\eta,\phi)
$$ 
are such that 
$$
\phi = \theta + f(t,x,\theta)
$$
and fulfill the inequalities (\ref{first-ine}), (\ref{second-ine}) and (\ref{third-ine}) for some constants $K_1(t), K_2(T)$ and $E(t)$. Using the uniform bound proved in Lemma \ref{stime-f11}:
$$
\| f(t,x,\theta)(\cdot) \|_{L^2} \le C_{00}(T) 
$$
we easily establish the existence of  the new constants $\tilde{K}_1 (t)$, $\tilde{K}_2 (T)$ and $\tilde{E}(t)$.
\endproof
\begin{theorem}
\label{isol}
Let us suppose $V(x) = \frac{1}{2}|x|^2 + V_0 (x)$ with   $\ds \sup_{x\in \Bbb R^n} \| \nabla^2 V_0 (x) \|<1$,  $(t,x,\eta) \in ]0,T] \times \Bbb R^{2n}$ and  $t \neq (2\tau +1)\frac{\pi}{2}$, $\tau\in \mathbb{N}$.  Then the following quadratic form:
$$
\langle \nabla^2_\theta S (t,x,\eta,\theta)  u , u \rangle,  \quad  \quad  u  \in \Bbb R^k,  
$$
is non degenerate on all points solving $\nabla_\theta S  (t,x,\eta,\theta) = 0$.
\end{theorem}
\proof
The quadratic form is non degenerate if and only if the solutions $\theta$ of the equation
\begin{eqnarray}
\label{isol-staz}
 \nabla_\theta S  (t,x,\eta,\theta) = 0
\end{eqnarray}
are isolated points. This  property can be translated in the infinite dimensional setting of the equation
$$
\frac{D \mathcal{S}}{D \phi} (t,x,\eta,\phi) = 0
$$
thanks to  the equivalence $\phi=\theta+f(t,x,\theta)$ shown in Lemma \ref{corr-sol}. We recall that
\vskip 4pt\noindent
\begin{eqnarray}
\mathcal{S}   =  \langle x , \eta \rangle +       \int_0^t  \langle  \int_0^s \phi^p (\tau) d\tau ,     \phi^x (s) \rangle  - \frac{1}{2m} \left(\eta +  \int_0^s \phi^p (\tau) d\tau\right)^2  -   V \left( x  -  \int_s^t  \phi^x (\tau) d\tau \right)  ds
\nonumber
\end{eqnarray}
\vskip 4pt\noindent
Now we perform the partial reduction of the infinite dimesional parameters, by means of the first stationarity equation $\displaystyle \frac{D \mathcal{S}}{D \phi^x} (t,x,\eta,\phi) = 0$ corresponding to $\ds m\phi^x (s) = \eta +  \int_0^s \phi^p (\tau) d\tau$ (essentially, the Legendre transform).  Therefore we get  the new functional:
\begin{eqnarray}
\widetilde{\mathcal{S}}(t,x,\eta,\phi^x)   =  \langle x -  \int_0^t \phi^x (\tau) d\tau, \eta \rangle +       \int_0^t   \frac{m}{2} \left[\left| \phi^x (s) \right|^2  -   V \left( x  -  \int_s^t  \phi^x (\tau) d\tau \right) \right] ds
\nonumber
\end{eqnarray}
Setting  $\gamma^x (s) = x  -  \int_s^t  \phi^x (\tau) d\tau$, we can consider the equivalent form
\begin{eqnarray}
\mathcal{A} [\gamma^x]    =  \langle \gamma^x (0), \eta \rangle +       \int_0^t   \frac{m}{2}\left[ \left| \dot{\gamma}^x  (s) \right|^2  -   V \left( \gamma^x (s) \right) \right] ds
\nonumber
\end{eqnarray}
with the boundary conditions: $\gamma^x (t)=x$ and $m\dot{\gamma}^x (0)= \eta$. The second variation is:
$$
\frac{D^2 \mathcal{A}}{D \gamma}(\gamma^x)[\delta \gamma,\delta \dot{\gamma} ]  =   \frac{1}{2} \int_0^t   m \left[| \delta \dot{\gamma}(s) |^2      - \nabla^2 V(\gamma^x (s)) \delta \gamma(s) \delta \gamma(s)    \right]\ ds
$$
\vskip 3pt\noindent
Writing down the integrand under the form
\vskip 8pt\noindent
\begin{equation*}
\Big(  \delta \gamma(s)  \  \delta \dot{\gamma}(s)   \Big) \ \left( \begin{array}{cc} - \nabla^2 V(\gamma^x(s)) &  0 \\  0 & m I  \end{array} \right) \  \left( \begin{array}{cc} \delta \gamma(s)  \\ \delta \dot{\gamma}(s) \end{array} \right) 
\end{equation*}
\vskip 4pt\noindent
we realize that requiring $\nabla^2 V(x)$ non-degenerate  $\forall \,x \in \Bbb R^n$, then the second variation is a bilinear non degenerate functional.
This implies that all the stationary curves of the action functional, namely the curves solving
$$
\frac{D \mathcal{A}}{D \gamma} (\gamma^x) [v] = 0  \quad \forall v \in T\Gamma
$$
are isolated  points belonging to $H^1([0,t];\Bbb R^n)$. We conclude that the same property must hold for the points $\theta \in \Bbb R^k$ solving equation (\ref{isol-staz}).
\endproof

Next,  we investigate  the number  of  solutions of the stationarity equation.
\begin{theorem}
\label{Th-locS}
Let us suppose $V(x) = \frac{1}{2}|x|^2 + V_0 (x)$ with   $\ds \sup_{x\in \Bbb R^n} \| \nabla^2 V_0 (x) \|<1$,  $(t,x,\eta) \in ]0,T] \times \Bbb R^{2n}$ and  $t \neq (2\tau +1)\frac{\pi}{2}$, $\tau\in \mathbb{N}$. Then the stationarity equation
$$
\nabla_\theta   S(t,x,\eta,\theta) =  0
$$
has a finite number of solutions $\theta^\star_\alpha(t,x,\eta)$, $1\le \alpha \le \mathcal{N}(t)$. The upper bound has the expression:
\begin{equation}
\label{eq-loc-N}
\mathcal{N}(t)  \le  \frac{ (2\tilde{E}(t))^k }{ \varepsilon(T)^k }
\end{equation}
Here $\tilde{E}(t)$ as in Theorem  \ref{compact} whereas 
\begin{equation}
\label{eps}
\varepsilon(T):= \frac{1}{{k}}\frac{ \inf_{(t,x,\eta,\theta)}  \sup_{i,j}  \left| \frac{\partial^2 S}{\partial \theta_i \partial \theta_j } \right| (t,x,\eta,\theta)}{ \sup_{(t,x,\eta,\theta)} \  \sup_{i,j,m} \left|\frac{\partial^3 S}{\partial \theta_i \partial \theta_j \partial \theta_m }  \right| (t,x,\eta,\theta)  + 1}.
 \end{equation}
 \end{theorem}
 \vskip 4pt\noindent
\proof
By Theorem (\ref{compact})  all critical parameters must be contained in the compact set $\overline{B}_r \subset \Bbb R^k$ with $r:=2\tilde{E}(t)$. As a consequence,  there exists a subsequence $\{\theta_{\alpha(j)} \}_{j \in \Bbb N}$ converging to some point $\bar{\theta}$ in $\overline{B}_r(0)$. However the function $\nabla_\theta   S(t,x,\eta,\cdot)$ is continuous on $\Bbb R^k$. Hence the limit is also a critical point, namely $0 = \nabla_\theta   S(t,x,\eta,\bar{\theta})$.  By the previous theorem all the critical points of $S$ are isolated. This is a contradiction, so their number must be finite. In order to obtain an upper bound for this number, we first observe that
\begin{eqnarray}
\nabla^2_\theta S (t,x,\eta,\theta) &=& \nabla^2_\theta S (t,x,\eta,\theta^\star)  + \int_0^1 \frac{d}{d\lambda} \nabla^2_\theta S (t,x,\eta,\theta^\star + \lambda (\theta-\theta^\star)) \ d \lambda
\nonumber\\
&=& \nabla^2_\theta S (t,x,\eta,\theta^\star)  + \int_0^1 D_\theta \nabla^2_\theta S (t,x,\eta,\theta^\star + \lambda (\theta-\theta^\star)) \ d \lambda \  ( \theta-\theta^\star )
\nonumber
\end{eqnarray}
We know that, thanks to Theorem \ref{isol},  the first  matrix on the righthand side is  non dege\-nerate. In order to verify that the addition of the second one does not change this property, we establish the matrix norm inequality:
\begin{equation}
\label{ineq1}
\left\| \int_0^1 D_\theta \nabla^2_\theta S (t,x,\eta,\theta^\star + \lambda (\theta-\theta^\star)) \ d \lambda \  ( \theta-\theta^\star ) \right\|_2 <  \| \nabla^2_\theta S (t,x,\eta,\theta^\star)\|_2.
\end{equation}
Here $\| \cdot \|_2$ is the usual norm for the matrix viewed as an operator.
Now denote $\varepsilon:= \| \theta - \theta^\star  \|$.  The above inequality is a fortiori verified if:
 \begin{equation}
\label{ineq2}
\varepsilon \ \sqrt{k}   \ \sup_{(t,x,\eta,\theta)} \  \sup_{i,j,m} \left|\frac{\partial^3 S}{\partial \theta_i \partial \theta_j \partial \theta_m }  \right| (t,x,\eta,\theta) <  \frac{1}{\sqrt{k}} \inf_{(t,x,\eta,\theta)}  \sup_{i,j}  \left| \frac{\partial^2 S}{\partial \theta_i \partial \theta_j } \right| (t,x,\eta,\theta)
\end{equation}
because the l.h.s is an upper bound for the l.h.s. of (\ref{ineq1}) and the r.h.s a lower bound for the r.h.s. of (\ref{ineq1}). 
(\ref{ineq2}) is in turn a fortiori verified if:
\begin{equation}
\label{ineq3}
\varepsilon \ \sqrt{k}   \left(  \sup_{(t,x,\eta,\theta)} \  \sup_{i,j,m} \left|\frac{\partial^3 S}{\partial \theta_i \partial \theta_j \partial \theta_m }  \right| (t,x,\eta,\theta)  + 1 \right) <  \frac{1}{\sqrt{k}} \inf_{(t,x,\eta,\theta)}  \sup_{i,j}  \left| \frac{\partial^2 S}{\partial \theta_i \partial \theta_j } \right| (t,x,\eta,\theta) 
\end{equation}
and this yields (\ref{eps}). In this way, we have found the radius $\varepsilon(T)$ of the balls in $\Bbb R^k$,  where each $\theta^\star$ is  a unique local critical point.   This local confinement of critical points together with the global one proved in Th \ref{compact}, allows us to get an estimate of their total number $N$. We simply compute the ratio between the volume of the ball $B_r$ containing all the points and the volume of the small isolating balls.
$$
\mathcal{N} (t) = \frac{vol(B_r)}{vol(B_\varepsilon)} = \frac{(2\tilde{E}(t))^k}{\varepsilon(T)^k} 
$$
\endproof

We use Theorem \ref{Th-locS} in order to study the global behaviour of the stationarity equation.
\begin{theorem}
\label{th-grad11}
Let us suppose $V(x) = \frac{1}{2}|x|^2 + V_0 (x)$ with   $\ds \sup_{x\in \Bbb R^n} \| \nabla^2 V_0 (x) \|<1$, $(t,x,\eta) \in ]0,T] \times \Bbb R^{2n}$ and  $t \neq (2\tau +1)\frac{\pi}{2}$, $\tau\in \mathbb{N}$. Let the number $\mathcal{N}(t)$ be given by (\ref{eq-loc-N}).  Then there exists a finite open partition $\ds \Bbb R^{2n} = \bigcup_{\ell=1}^{\mathcal{N}(t)} D_\ell$ such that  the equation 
\begin{equation}
\label{eq-grad11}
0= \nabla_\theta   S(t,x,\eta,\theta) 
\end{equation}
admits  on each $D_\ell$  exactly $\ell$ smooth solutions $\theta^\star_\alpha (t,x,\eta)$, $1\le  \alpha \le \ell$ . 
\end{theorem}
\proof  We recall that $\Sigma_S := \{ (x,\eta,\theta) \in \Bbb R^{2n+k}  | \  0 = \nabla_\theta S (t,x,\eta,\theta)  \}$ is a $2n$-dimensional submanifold of $\Bbb R^{2n+k}$ diffeomorphic to $\Lambda_t$. Moreover, by the nondegeneracy hypothesis on $\nabla^2 V$ we have the transversal behaviour of $\Sigma_S$ (with respect to $(x,\eta) \in \Bbb R^{2n}$) almost everywhere;  namely the rank of $\nabla^2_\theta S$ can differ from its maximum value ($k$) only on subsets whose projection on $(x,\eta) \in \Bbb R^{2n}$ is of zero measure. The condition of transversality is fulfilled on components $D_\ell$ (locally diffeomorphic to open sets of $\Bbb R^{2n}$) where  the local smooth inversion of equation $(\ref{eq-grad11})$ is possible,  yielding $\ell$ functions   $\theta^\star_\alpha (t,x,\eta)$. This argument works up to the finite maximum value $\mathcal{N}(t)$.
\endproof

\bigskip
\bigskip

\subsection{Transport equations}
We conclude this section by introducing transport equations in a global geometrical setting.

\begin{theorem}
Let us consider $\rho \in \mathcal{S}(\Bbb R^k;\Bbb R)$ with  $\| \rho \|_{L^1} =1$.  
The transport equation written on the stationary points $\Sigma_S$ of the generating function $S$, 
\begin{equation}
\label{eq-tr1} \left\{
\begin{array}{l}
\displaystyle{ \partial_t b_0 + \frac{1}{m} \nabla_x S \ \nabla_x b_0  + \frac{1}{2m} \Delta_x S  \ b_{0}   (t,x,\eta,\theta) = 0, }
\\
\\
b_0  ( 0,x,\eta,\theta) = \rho(\theta), \quad (x,\eta,\theta) \in \Sigma_S .
\end{array}
\right.
\end{equation}
admits the following solution:
\begin{equation}
\label{simb0}
b_{0} (t,x,\eta,\theta)   =  \exp\left\{ - \frac{1}{2m} \int_0^t \Delta_x  S( \tau , \gamma^x (t,x,\theta)(\tau) ,\eta,\theta)  d\tau \right\}  \rho(\theta)
\end{equation}
where $\gamma^x $ is the family of curves defined in (\ref{fin-curv}).
\end{theorem}
\proof
The inital condition is immediately verified: 
$$
b_0  (0,x,\eta,\theta) = \rho (\theta)
$$
Recalling the results of  Proposition \ref{prop-sv0} and Theorem \ref{prop-sv10}, we compute the expression of the   differential operator $\ds  \partial_t b_0 + \frac{1}{m} \nabla_x S  \ \nabla_x b_0 (t,x,\eta,\theta)$ when evaluated on the submanifold $\Sigma_S := \{ (x,\eta,\theta) \in \Bbb R^{2n+k} \ | \ 0 =  \nabla_\theta  S  (t,x,\eta,\theta) \}$. Namely,
\begin{eqnarray}
\partial_t b_0 + \frac{1}{m} \nabla_x S  \ \nabla_x b_0 (t,x,\eta,\theta)  \Big|_{\Sigma_S}  &=&  \partial_t b_0 + \frac{1}{m} \left( \eta + \gamma^p (t,x,\theta)(t) \right)  \nabla_x b_0 (t,x,\eta,\theta) \Big|_{\Sigma_S}    
\nonumber\\
&=&  \partial_t b_0 +  \dot{\gamma}^x (t,x,\theta)(t)   \nabla_x b_0 (t,x,\eta,\theta) \Big|_{\Sigma_S}    
\nonumber\\
&=& \partial_\mu b_0 (\mu,x,\eta,\theta) +  \dot{\gamma}^x (t,x,\theta)(\mu)   \nabla_x b_0 (\mu,x,\eta,\theta) \Big|_{\Sigma_S}  \Big|_{\mu=t}  
\nonumber\\
&=& \frac{d}{d\mu} b_0 (\mu, \gamma^x (t,x,\theta)(\mu),\eta,\theta) \Big|_{\Sigma_S}  \Big|_{\mu=t}  
\end{eqnarray}
where the expression of $\ds \frac{1}{2m} \Delta_x S  (t,x,\eta,\theta)   \ b_{0}   (t,x,\eta,\theta) $ is:
$$
\frac{1}{2m} \Delta_x S (t,x,\eta,\theta)  \ b_{0}   (t,x,\eta,\theta) = \frac{1}{2m} \Delta_x S(t,\gamma^x (t,x,\theta)(\mu),\eta,\theta)  \ b_{0}   (\mu,\gamma^x (t,x,\theta)(\mu),\eta,\theta)  \Big|_{\mu=t}    
$$
Now, we write down the equation in the new variable $\mu$ and for all $(x,\eta,\theta) \in \Bbb R^{2n+k}$:
$$
\frac{d}{d\mu} b_0 (\mu, \gamma^x (t,x,\theta)(\mu),\eta,\theta) + \frac{1}{2} \Delta_x S (\mu,\gamma^x (t,x,\theta)(\mu),\eta,\theta)  \ b_{0}   (\mu,\gamma^x (t,x,\theta)(\mu),\eta,\theta) = 0
$$
If  we define $\alpha(\mu):= b_0 (\mu, \gamma^x (t,x,\theta)(\mu),\eta,\theta)$ we can rewrite the previous equation as
$$
\frac{d}{d\mu}  \alpha(\mu)  =  - \frac{1}{2m} \Delta_x S (\mu,\gamma^x (t,x,\theta)(\mu),\eta,\theta)  \ \alpha(\mu).
$$
where  the variables $(t,x,\eta,\theta)$ have to be considered as fixed. This yields:
$$
b_0 (\mu, \gamma^x (t,x,\theta)(\mu),\eta,\theta)  =   \exp\left\{ - \frac{1}{2m} \int_0^\mu \Delta_x  S( \tau , \gamma^x (t,x,\theta)(\tau) ,\eta,\theta)  d \tau \right\}   \rho(\theta)
$$
Finally, we make $\mu=t$ and so we obtain the solution of the original problem (\ref{eq-tr1}):
$$
b_{0} (t,x,\eta,\theta)   =   \exp\left\{ - \frac{1}{2m} \int_0^t \Delta_x  S( \tau , \gamma^x (t,x,\theta)(\tau) ,\eta,\theta)  d \tau \right\}   \rho(\theta)
$$
\endproof

\begin{theorem}
\label{Th-b_00}
Let  $b_0$ be defined as in  (\ref{simb0}) with $\rho(\theta):= e^{-|\theta|^2} \xi(\theta)$ and $\xi \in  C^\infty_b (\Bbb R^k;\Bbb R^+)$. Then $b_0 (t,x,\eta,\theta) \in C^\infty([0,T] \times \Bbb R^{2n+k}; \Bbb R^+)$ and $b_0 (t,x,\eta,\cdot) \in \mathcal{S} (\Bbb R^k;\Bbb R^+)$ for every $(t,x,\eta)$ fixed.
Moreover,  there exists a constant $C^+ (T)>0$  such that 
\begin{eqnarray}
| \partial_x^\alpha b_0 (t,x,\eta, \theta )| &\le&  C_\alpha^+ (T) e^{d_{\alpha}(T) \lambda(x,\eta)}  e^{-|\theta|^2}  \quad \forall (x,\eta,\theta) \in \Bbb R^{2n+k}
\label{stime-b00}
\end{eqnarray}
\end{theorem}
\proof
Let us first obtain a more explicit expression for  $(\Delta_x)S(\cdot)$:
\begin{eqnarray}
\Delta_x  S( t , x ,\eta,\theta)  &=&   2tr(L)t \  + \langle \Delta_x  \nu(t,x,\theta) , \theta \rangle + \langle v(t,x,\eta) , \Delta_x f(t,x,\theta) \rangle 
\nonumber\\
&+&  2 \langle \nabla_x v(t,x,\eta) , \nabla_x f(t,x,\theta) \rangle +   \Delta_x g (t,x,\theta)
\nonumber\\
&=&   2 tr(L)t \ + \langle 2R(t) \Delta_x  f(t,x,\theta) , \theta \rangle + \langle v(t,x,\eta) , \Delta_x f(t,x,\theta) \rangle
\nonumber\\
&+&  2 \langle \nabla_x v(t,x,\eta) , \nabla_x f(t,x,\theta) \rangle +  \Delta_x g (t,x,\theta)
\nonumber\\
&=&  2 tr(L)t \ + \langle 2R(t) \Delta_x  f(t,x,\theta) , \theta \rangle + \langle v(t,x,\eta) , \Delta_x f(t,x,\theta) \rangle 
\nonumber\\
&+&  2 \langle \nabla_x v(t,x,\eta) , \nabla_x f(t,x,\theta) \rangle + \Delta_x g (t,x,\theta) 
\end{eqnarray}
Now, we recall that
$$
\gamma^x (t,x,\theta)(\tau) = x - \int_\tau^t \theta^x (r) + f^x (t,x,\theta^x)(r) \ dr
$$
where $f$ and all its derivatives are $L^2$ uniformly bounded, as proved in Lemma  \ref{stime-f11}, whereas $v$ is linear in $(x,\eta)$ and $g$ is $L^\infty$ bounded. Now we observe that by setting $\rho(\theta):= e^{-|\theta|^2} \xi(\theta)$ with a bounded  $\xi \in  C^\infty (\Bbb R^k;\Bbb R^+)$  then $b_0 (t,x,\eta,\cdot)$  is a Schwartz function on $\Bbb R^k$. Indeed,
$$
| b_0 (t,x,\eta,\theta) | \le   \exp\Big\{ \frac{1}{2m}  \int_0^t  |\Delta_x  S( \tau , \gamma^x (t,x,\theta)(\tau) ,\eta,\theta) | d\tau   \Big\} e^{-|\theta|^2} \xi(\theta)
$$
But by the above detailed computation we see that
\begin{eqnarray}
\lefteqn{ |\Delta_x  S( \tau , \gamma^x(t,x,\theta) ,\eta,\theta) |  }
\nonumber\\
&\le& | 2{\rm tr}(L)t| + 2 \| R(t) \| \ \|  \Delta_x  f(t,\gamma^x ,\theta) \|_{L^2}  \| \theta \|  +  \| v(t,\gamma^x,\eta) \|_{L^2}  \|  \Delta_x  f(t,\gamma^x,\theta) \|_{L^2}
\nonumber\\
&+&    2 \|  \nabla_x  v(t,\gamma^x,\eta) \|_{L^2}  \| \nabla_x f(t,\gamma^x,\theta) \|_{L^2} +  \| \Delta_x g (t,\gamma^x,\theta)\|_{L^\infty}
\nonumber
\end{eqnarray}
$\|  \Delta_x  f(t,\gamma^x(t,x,\theta),\theta) \|_{L^2}  \| \theta \|$ is linear in $\theta$ and  $\| v(t,\gamma^x(t,x,\theta),\eta) \|_{L^2}  \|  \Delta_x  f(t,\gamma^x(t,x,\theta),\theta) \|_{L^2}$ has  a linear  uniform growth on $(x,\eta,\theta)$. To see this, remark that
$$
\| v(t,\gamma^x(t,x,\theta),\eta) \|_{L^2}  \le \| v(t,x,\eta) \|_{L^2} + \| v(t, \int_\tau^t \theta^x (r)  dr,\eta) \|_{L^2}  + \| v(t,\int_\tau^t  f^x (t,x,\theta^x)(r)  dr,\eta) \|_{L^2}  
$$
The first  and third term on the right hand side generate a linear growth on $(x,\eta)$, the second term  has a linear dependence on $\theta$. The other terms above are bounded with respect to all variables. We conclude that  $b_0$ has a uniform exponential behaviour on all its variables, and that the function $\rho(\theta):= e^{-|\theta|^2} \| \xi \|_{C^0}$ makes the effective dependence on $\theta$ of Schwartz type.
\endproof

\begin{theorem}
\label{geo-tran2}
Let  $S$ and $\Sigma_S$ be as in Theorem \ref{eq-hj-r}. Then there exists $\widetilde{\Theta}_N \in C^\infty_b ( [0,T]  \times \Bbb R^{2n+k} ; \Bbb R )$ with $\widetilde{\Theta}_N |_{\Sigma_S}=0$, such that the solution $S_N$   of the
problem
\begin{equation}
\label{eq-tr12} \left\{
\begin{array}{l}
\displaystyle{ \partial_t b_{0,N} + \frac{1}{m} \nabla_x S_N \ \nabla_x  b_{0,N}  + \frac{1}{2m} \Delta_x  S_N  \ b_{0,N}   (t,x,\eta,\theta)  =  \widetilde{\Theta}_N, }
\\
\\
b_{0,N}  ( 0,x,\eta,\theta) = \rho(\theta).
\end{array}
\right.
\end{equation}
 enjoys the property:
 \begin{eqnarray}
\label{reg-div2}
\Pi^{ j } ( \widetilde{\Theta}_N )  \in C^\infty_b ( [0,T]  \times \Bbb R^{2n+k} ; \Bbb R ) \quad  \quad \forall 1 \le j \le N, \quad N=1,2,\ldots
\end{eqnarray}
where, as in Theorem \ref{theo-S-eq}:
\begin{eqnarray*}
\Pi ( \widetilde{\Theta}_N ) :=   {\rm div}_\theta \left( \widetilde{\Theta}_N \frac{ \nabla_\theta S_N }{|\nabla_\theta S_N |^2} \right)
\end{eqnarray*}
\end{theorem}
\proof 
Let us define 
\begin{eqnarray}
\label{b0N}
b_{0,N} (t,x,\eta,\theta)   :=   \exp\left\{ - \frac{1}{2m} \int_0^t \Delta_x  S_N ( \tau , \gamma^x (t,x,\theta)(\tau) ,\eta,\theta)  d \tau \right\}   \rho(\theta)
\end{eqnarray}
and prove that is solves the above problem. Indeed, we can write down the expantions for $z:=(x,\eta,\theta)$ around $\bar{z} \in \Sigma_{S_N} = \Sigma_S$ 
\begin{eqnarray}
\label{b0N-def}
b_{0,N} (t,z) &=& b_{0} (t,\bar{z}) + f_N(t,z)
\\
\nabla_x b_{0,N} (t,z) &=& \nabla_x b_{0} (t,\bar{z}) + g_N(t,z)
\\
\partial_t b_{0,N} (t,z) &=& \partial_t b_{0} (t,\bar{z}) + h_N(t,z)
\end{eqnarray}
all these terms are related to the choice of $G_N$ in Theorem \ref{theo-S-eq} and  their rate of convergence to zero near $\bar{z}$ are related as well. By unperturbed equation (\ref{eq-tr1}), we compute:
\begin{eqnarray}
\widetilde{\Theta}_N (t,z)  &=&\partial_t b_{0,N} + \frac{1}{m} \nabla_x S_N \ \nabla_x  b_{0,N}  + \frac{1}{2m} \Delta_x  S_N  \ b_{0,N}   (t,z)  
\nonumber\\
&=& \partial_t b_{0} (t,\bar{z}) + h_N (t,z) +  \frac{1}{m} \left(  \nabla_x S (t,\bar{z}) + F_N^x (t,z)  \right) \left(  \nabla_x b_{0} (t,\bar{z}) + g_N (t,z)  \right) 
\nonumber\\
&+&  \frac{1}{2m} ( \Delta_x  S(t,\bar{z}) + G_N (t,z) )  (  b_{0} (t,\bar{z}) + f_N (t,z)  )
\nonumber\\
&=& h_N (t,z) +  \frac{1}{m}  \nabla_x S (t,\bar{z}) g_N (t,z)  +  F_N^x (t,z) \left(  \nabla_x b_{0} (t,\bar{z}) + g_N (t,z)  \right)
\nonumber\\
&+&  \frac{1}{2m}  \Delta_x  S(t,\bar{z}) f_N (t,z) + \frac{1}{2m}  G_N (t,z)   ( b_{0} (t,\bar{z}) + f_N (t,z) )
\end{eqnarray}
Moreover, by Theorem \ref{theo-S-eq}, we remember that around $\bar{z} \in \Sigma_{S_N} = \Sigma_S$ the new stationarity equation is
$$
\nabla_\theta  S_N (t,z) = F_N^\theta (t,z)
$$
Now we can state that a suitable choice of $F^\theta_N,F^x_N$ and $G_N$ leads to the following property:
$$
\Pi^{j} (\widetilde{\Theta}_N)  \in C^\infty_b ( [0,T]  \times \Bbb R^{2n+k} ; \Bbb R ) \quad  \quad \forall 1 \le j \le N
$$
\endproof

\section{A class of global FIO}
\renewcommand{\theequation}{\thesection.\arabic{equation}}
\setcounter{equation}{0}%
In this section we follow the general setting of H\"ormander \cite{Ho}, and in particular we study  a class of  FIO related to the Hamiltonian flow $\phi^t_{\mathcal{H}}$,  by using the generating functions constructed in the previous section.  The study of the topology of their critical points  will be useful to determine important analitical properties of the FIO such as asymptotic behaviour of the kernel and $L^2$-continuity.

\subsection{Basic definition and main properties}
First, we introduce the set of phase functions:
\begin{definition}
\label{PF}
The set of phase functions $S(t,x,\eta,\theta): [0,T] \times \Bbb R^n \times \Bbb R^n \times \Bbb R^k \rightarrow \Bbb R$ is the set of smooth global generating functions of the graphs $\Lambda_t \subset T^\star \Bbb R^n \times T^\star \Bbb R^n$ of the  canonical maps  $\phi^t_H :T^\star \Bbb R^n \rightarrow T^\star \Bbb R^n$, with the inititial condition $S(0,x,\eta,\theta) = \langle x ,\eta \rangle$. Each $\Lambda_t$ admits the parametrization:
\begin{eqnarray}
\Lambda_t &:=& \left\{ (y,\eta;x,p) \in T^\star \Bbb R^n \times T^\star \Bbb R^n \ | \  (x,p) = \phi_\mathcal{H}^t (y,\eta) \right\}  
\nonumber\\
&=& \left\{ (y,\eta;x,p) \in T^\star \Bbb R^n \times T^\star \Bbb R^n \ | \  p = \nabla_x S, \quad y =  \nabla_\eta S, \quad   0 =  \nabla_\theta S (t, x,\eta,\theta)  \right\}  
\nonumber
\end{eqnarray}
\end{definition}
Before going further, we recall that by Theorem \ref{compact}, the generating function $S$ enjoys an important property. Namely, consider the set of critical points
\begin{equation}
\label{crset}
\Sigma_S  :=  \{  (x,\eta,\theta)  \in \Bbb R^{2n+k} \ | \  0  =  \nabla_\theta S(t,x,\eta,\theta)  \}.
\end{equation}
Then $\Sigma_S $ is a manifold globally diffeomorphic to $\Lambda_t$;  moreover for all $t>0$ the following set 
\begin{equation}
\label{Omega2}
\Upsilon_S  :=  \{  (x,\eta,\theta)  \in \Bbb R^{2n+k} \ | \  |x|^2 + |\eta|^2 > D(T)^2,  \  |\theta|  \le  \tilde{K}_2(T) \lambda(x,\eta)   \} 
\end{equation}
is free from critical points, i.e.:
$$
\Upsilon_S   \subset \Bbb R^{2n+k}  \backslash  \Sigma_S
$$ 
Second,  we introduce the relevant class of symbols associated to $S$:
\begin{definition}
\label{def-simb}
The set of symbols  consists of all $b \in C^{\infty}([0,T] \times \mathbb{R}^{2n} \times \mathbb{R}^k; \mathbb{R})$ such that 
\begin{itemize}
\item[(i)]
$$
 b(0,x,\eta,\theta) = \rho(\theta), \quad  \rho (\cdot)  \in \mathcal{S}(\Bbb R^k;\Bbb R^+), \quad  \int_{\Bbb R^k} \rho(\theta) d\theta = 1.
 $$ 
\item[(ii)] For all  multi-indices $\alpha, \beta, \sigma$  and $t \in ]0,T]$ the inequalities 
\begin{equation}
\label{prop-simb}
| b(t,x,\eta,\theta) | \le 
\left\{
\begin{array}{l}
\displaystyle{ C^{+} (T) \  e^{\lambda(x,\eta)}
 e^{-|\theta|^2}  ,    \quad \quad  (x,\eta,\theta) \notin   \Upsilon_S }
\\
\\
\displaystyle{ C^{-} (T)  \lambda^{-n} (x,\eta)    e^{-|\theta|^2}
,    \quad (x,\eta,\theta) \in  \Upsilon_S  }
\end{array}
\right.
\end{equation}
hold  for some  constants $C^{\pm}_{\alpha,\beta,\sigma}(T)>0$.
\end{itemize}
\end{definition}

\begin{remark}
\label{Remark-simb}
{\rm The exponential upper bound outside $\Upsilon_S$   is verified by the symbol $b_0$ (see Th. \ref{Th-b_00}) and also, as we will see, by any other symbol $b_j, j=1,\ldots$ entering in Theorem 1.1.  Moreover, on domain $\Upsilon_S$  there are no critical points for the function $S$ and  this leads to require asymptotic vanishing behaviour  of type $\lambda^{-n}(x,\eta) e^{-|\theta|^2}$ in this region for $b_0$; as a  conseguence the same asymptotic property is fulfilled by all $b_j$. This setting is motivated by the fact that the contribution of this region to the FIO can be of order  $O(\hbar^\infty)$ as we see in Corollary \ref{FIO-Y}. In this framework, we provide a very simple proof of global  $L^2$ continuity.}
\end{remark}

Finally, we introduce the  class of global FIO associated to the Hamiltonian flow $\phi_\mathcal{H}^t$:
\begin{definition}
\label{def-B}
Fix a phase function $S$ as in Definition \ref{PF}, and a symbol $b$ as in Definition \ref{def-simb}. Then the global $\hbar$-Fourier Integral Operator on $\mathcal{S} (\Bbb R^n)$ is defined as:
\begin{equation}
\label{first-FIO11}
B(t) \varphi (x)  =   (2\pi \hbar)^{-n}   \int_{\Bbb R^n} \int_{\Bbb R^n} \int_{\Bbb R^k}   e^{\frac{i}{\hbar} ( S(t,x,\eta,\theta)  - \langle y , \eta \rangle )   } \  b (t,x,\eta,\theta) \ d\theta \ d\eta \ \varphi(y) \ dy 
\end{equation}
In equivalent way, it can be rewritten in the form:
\begin{equation}
B(t) \varphi (x)  =    (2\pi \hbar)^{-n} \int_{\Bbb R^n} \int_{\Bbb R^k}   e^{ \frac{i}{\hbar} \widetilde{S}(t,x,y,u)   } \  \widetilde{b} (x,u) \ du \  \varphi(y) \ dy 
\end{equation}
where $u:=(\eta,\theta)$, $\widetilde{S}(t,x,y,u) := S(t,x,\eta,\theta) - \langle y,\eta \rangle$ and $\widetilde{b} (t,x,u)  := b(t,x,\eta,\theta)$. Indeed, if $S$ generates the Lagrangian submanifold $\Lambda$, then $\widetilde{S}$ does the same in new variables:
$$
\Lambda   =   \{  (x,p;y,\eta)  \in  T^\star \Bbb R^n \times T^\star \Bbb R^n  \ | \ p = \nabla_x \widetilde{S}, \quad \eta = - \nabla_y \widetilde{S}, \quad 0 = \nabla_u \widetilde{S}    \}
$$
\end{definition}

\begin{theorem}
\label{bound-proof}
Let us consider the FIO as in Definition  \ref{def-B}. 
$$
B(t) \varphi (x)  =  (2\pi \hbar)^{-n}  \int_{\Bbb R^n} \int_{\Bbb R^n} \int_{\Bbb R^k}   e^{\frac{i}{\hbar} ( S(t,x,\eta,\theta)  - \langle y , \eta \rangle )   } \  b (t,x,\eta,\theta) \ d\theta \ d\eta \ \varphi(y) \ dy 
$$
Then $B(t): \mathcal{S}(\Bbb R^n)\to \mathcal{S}(\Bbb R^n)$ is continuous  and admits a continuous extension as an operator in  $L^2(\Bbb R^n)$.
\end{theorem}
\proof
We begin by rewriting the FIO under the form of an integral operator acting on the $\hbar$-Fourier transform of the initial datum: 
\begin{eqnarray}
B(t) \varphi (x)  &=&  (2\pi \hbar)^{-n}   \int_{\Bbb R^n} \int_{\Bbb R^k}   e^{\frac{i}{\hbar}  S(t,x,\eta,\theta)    } \  b (t,x,\eta,\theta) \ d\theta \hat{\varphi}_\hbar(\eta)  d\eta
\nonumber\\
&=& (2\pi \hbar)^{-n}  \int_{\Bbb R^n}  \widehat{\sigma}_\hbar  (t,x,\eta)  \hat{\varphi}_\hbar (\eta)  d\eta. 
\\
\nonumber
\widehat{\sigma}_\hbar  (t,x,\eta) & :=& \int_{\Bbb R^k}e^{\frac{i}{\hbar}  S(t,x,\eta,\theta)    } \  b (t,x,\eta,\theta) \ d\theta
\end{eqnarray}
This is because of the integral in the $\theta$-variables is absolutely convergent since $b(t,x,\eta,\cdot) \in \mathcal{S}(\Bbb R^k)$,  and $\varphi(y)$ is also a Schwartz function and therefore admits a $\hbar$-Fourier transform in $\mathcal{S}(\Bbb R^n)$. The absolute convergence of the integral, as well as the $L^2$-continuity, is the consequence of the following computations.
\begin{eqnarray}
\widehat{\sigma}_\hbar (t,x,\eta) &=& \int_{\Bbb R^k}   e^{ \frac{i}{\hbar}  S(t,x,\eta,\theta) } \  b (t,x,\eta,\theta) \ d\theta = e^{ \frac{i}{\hbar}  \langle x , \eta \rangle } [ \widehat{\sigma}_\hbar^+  (t,x,\eta)  +  \widehat{\sigma}_\hbar^-  (t,x,\eta)  ]
\\
\widehat{\sigma}_\hbar^-(t,x,\eta)   &=&   \int_{ B_\delta (0) \subset  \Bbb R^k }          e^{ \frac{i}{\hbar}  ( S(t,x,\eta,  \theta) - \langle x , \eta \rangle ) } \  b (t,x,\eta,  \theta) \ d\theta
\\
\widehat{\sigma}_\hbar^+  (t,x,\eta)  &=&   \int_{ \Bbb R^k \backslash B_\delta (0) }   e^{ \frac{i}{\hbar}  ( S(t,x,\eta,  \theta) - \langle x , \eta \rangle ) } \  b (t,x,\eta,  \theta) \ d\theta
\end{eqnarray}
with $\delta := \tilde{K}_2(T) \lambda(x,\eta)$. For $t=0$ we  have  $\widehat{\sigma}_\hbar^+  (0,x,\eta)  +  \widehat{\sigma}_\hbar^-  (0,x,\eta) = 1$, $B(0) \varphi= \varphi$,  and the continuity is obvious.  For $t>0$ we can apply the  estimates of Property (ii) of  Definition \ref{def-simb}. In the region containing the critical points we have: 
\begin{eqnarray}
| \widehat{\sigma}_\hbar^+   (t,x,\eta) |  & \le &  \int_{ \Bbb R^k \backslash B_\delta (0)  }   | b (t,x,\eta,  \theta) |  \  d\theta \le  \int_{ \Bbb R^k \backslash B_\delta (0)  }   C^{+}_{0}(T)  e^{\lambda(x,\eta)}  e^{-|\theta|^2}   d\theta
\nonumber\\
&=&  C^{+}_{0}(T)  e^{\lambda(x,\eta)}  \int_{ \Bbb R^k \backslash B_\delta (0)  }   e^{-|\theta|^2}    d\theta  
\end{eqnarray}
By writing down the integral in spherical coordinates, we have the following simple estimates
$$
\int_{ \Bbb R^k \backslash B_\delta (0)  }   e^{-|\theta|^2}   d\theta =  c_k   \int_\delta^\infty  e^{-\rho^2}   \rho^{k-1}  d\rho \le  c_k  d_k (L)  \int_\delta^\infty  e^{-\rho L}  d\rho =  c_k  d_k (L)    e^{- L \delta}  
$$
for all $L > 0$ and $d_k (L) := \sup_{\rho \ge 0} e^{-\rho^2}   \rho^{k-1} e^{\rho L}$. In particular we choose $L  := 1 + \tilde{K}_2^{-1}(T)$, so that it follows
\begin{eqnarray}
| \widehat{\sigma}_\hbar^+   (t,x,\eta) | \le C^{+}_{0}(T)  e^{\lambda(x,\eta)}  c_k d_k (L) e^{ - \tilde{K}_2(T) \lambda(x,\eta) - \lambda(x,\eta)} =  C^{+}_{0}(T)    c_k  d_k (L)  e^{ - \tilde{K}_2(T) \lambda(x,\eta)} 
\end{eqnarray}
Whereas in the other region we can write:
\begin{eqnarray}
| \widehat{\sigma}_\hbar^-  (t,x,\eta) |  & \le & \int_{ B_\delta (0) \subset  \Bbb R^k }   | b (t,x,\eta, \theta) |   d\theta  \le \int_{ B_{ \delta} (0) \subset  \Bbb R^k }   C^{-}_{0}(T)  \lambda^{-n}(x,\eta)  e^{-|\theta|^2}    d\theta 
\nonumber\\
& \le &   \int_{ \Bbb R^k }   C^{-}_{0}(T)  \lambda^{-n}(x,\eta)  e^{-|\theta|^2}    d\theta  
=  C^{-}_{0}(T) \pi^{-\frac{k}{2}}     \lambda^{-n}(x,\eta)   
\end{eqnarray}
We can now apply the Schur Lemma to both integral operators  and this yields the $L^2$-boundedness.  
\endproof

Now we prove a result,  based on an argument of Duistermaat (see \cite{Dui}, Prop 2.1.1).
\begin{lemma}
\label{F-staz-gen}  Let us consider a FIO of type (\ref{first-FIO11}) with  phase function $S$ and symbol $g$ leading to a convergent integral. Suppose that
\vskip 3pt\noindent
\begin{equation}
\label{Ldag}
\Pi ( g ) :=  {\rm div}_\theta \left( g  \frac{ \nabla_\theta S}{\| \nabla_\theta S \|^2}   \right) \in C^\infty([0,T]\times {\Bbb R^{2n}}\times {\Bbb R^k};{\Bbb R})
\end{equation}
\vskip 3pt\noindent
and that the FIO with symbol $\Pi(g)$ is  convergent.  Then,  the following equivalence  holds:
\vskip 10pt\noindent
\begin{equation}
\label{int-S}
\int_{\Bbb R^n} \int_{\Bbb R^k}   e^{\frac{i}{\hbar}  S(t,x,\eta,\theta)  }  g (t,x,\eta,\theta) \ d\theta \hat{\varphi}(\eta) d\eta  = - i \hbar \int_{\Bbb R^n}  \int_{\Bbb R^k}   e^{\frac{i}{\hbar}  S(t,x,\eta,\theta)   }  \Pi g(t,x,\eta,\theta) \ d\theta  \hat{\varphi}(\eta) d\eta
\end{equation}
\vskip 3pt\noindent
\end{lemma}
\proof 
The differential operator $\displaystyle \mathbb{L} \psi :=  \frac{ \langle \nabla_\theta S, \nabla_\theta \psi  \rangle}{\| \nabla_\theta S \|^2}$ verifies the relation
$$
- i \hbar    \mathbb{L}    e^{\frac{i}{\hbar}  S}   =   e^{\frac{i}{\hbar}  S}
$$
Now, by using integration by parts and  the definition of the operator $\mathbb{L}$, we get:
$$
\int_{\Bbb R^k}   e^{\frac{i}{\hbar}  S    }  g   d\theta   = - i \hbar \int_{\Bbb R^k}   \mathbb{L} \left( e^{\frac{i}{\hbar}  S   } \right) g   d\theta 
= -i   \hbar  \int_{\Bbb R^k}   e^{\frac{i}{\hbar}  S    }  \Pi ( g )   d\theta
$$
\endproof

\begin{corollary}
\label{FIO-Y}
Let $\tilde{g} \in C^\infty ([0,T] \times \Bbb R^{2n+k}; \Bbb R) $ be such that $\Pi^j (\tilde{g}) \in C^\infty ([0,T] \times \Bbb R^{2n+k}; \Bbb R)$ and   that  the corresponding FIO is convergent  for all $0 \le j \le N$. Then: 
$$
\int_{\Bbb R^n} \int_{\Bbb R^k}   e^{\frac{i}{\hbar}  S(t,x,\eta,\theta)  }  \tilde{g} (t,x,\eta,\theta) \ d\theta \hat{\varphi}(\eta) d\eta  = (- i \hbar)^N \int_{\Bbb R^n}  \int_{\Bbb R^k}   e^{\frac{i}{\hbar}  S(t,x,\eta,\theta)   }  \Pi^N \tilde{g} (t,x,\eta,\theta) \ d\theta  \hat{\varphi}(\eta) d\eta
$$ 
\end{corollary}
\proof
The iterated application of the previous Lemma gives the result.  \endproof

\begin{remark}
If we take two symbols $g_1, g_2$ coinciding on $\Sigma_S$ and moreover such that $g_1 - g_2 = \tilde{g}$, $\tilde{g}$ as in the above Corollary, then  the related FIO coincide up to order $O(\hbar^N)$. 
\end{remark}

\section{Global parametrices of the evolution operator}
\renewcommand{\theequation}{\thesection.\arabic{equation}}
\setcounter{equation}{0}
Here we prove the main result of this paper.  Consider  the  initial-value problem for Schr\"odinger equation
\begin{equation}
\label{eqSch99} \left\{
\begin{array}{l}
{\displaystyle i \hbar \partial_t \psi(t,x) =   - \frac{\hbar^2}{2m}  \Delta \psi(t,x) + V(x) \psi(t,x),}
\\
\\
\psi(0,x) = \varphi(x) \in \mathcal{S}(\Bbb R^n).
\end{array}
\right.
\end{equation}
with a potential $V$ quadratic at infinity,  of the type (\ref{H-def1}).\\
We proceed to apply the results of the previous two sections in order to prove Theorem 1.1; namely, to construct a parametrix for the evolution operator under the form of series of a global FIO such that  the solution of the Schr\"odinger equation (\ref{eqSch99}) admits the following representation:
\begin{equation}
\label{rep99}
\psi(t,x) =  \sum_{j=0}^\infty (2\pi\hbar)^{-n} \int_{\Bbb R^{2n+k}}    e^{\frac{i}{\hbar} ( S( t,x,\eta,\theta)  - \langle y , \eta \rangle )   }   \hbar^j b_{j} ( t,x,\eta,\theta) \ d\theta  \ d\eta  \  \varphi(y) \ dy + O(\hbar^\infty)
\nonumber
\end{equation}
within the time interval $t \in [0,T]$ with $T$ arbitrary large.
\vskip 6pt\noindent
{\bf Proof of Theorem 1.1}
\vskip 5pt\noindent
Denoting $\displaystyle H_x :=  - \frac{\hbar^2}{2m}  \Delta_x + V(x)$ the action of the Schr\"odinger operator we look for a family of global FIO $\{B_j (t)\}_{j\in \Bbb N}$ with symbol $b_j(t,x,\eta,\theta)$ enjoying Properties (i) and (ii) of Defi\-nition  \ref{def-B} such that
$$
0 = \left(  H_x  - i \hbar \partial_t  \right)  \sum_{j=0}^\infty \int_{\Bbb R^{2n+k}}    e^{\frac{i}{\hbar} ( S( t,x,\eta,\theta)  - \langle y , \eta \rangle )   }   \hbar^j b_{j} ( t,x,\eta,\theta) \ d\theta  \ d\eta  \  \varphi(y) \ dy + O(\hbar^\infty)
$$
First of all, the approximation of order  zero is the operator $B_0(t)$ defined as:
\begin{equation}
\label{def-semi}
B_0(t) \varphi := (2\pi\hbar)^{-n}  \int_{\Bbb R^{2n+k}}    e^{\frac{i}{\hbar} ( S( t,x,\eta,\theta)  - \langle y , \eta \rangle )   }  b_{0} ( t,x,\eta,\theta) \ d\theta  \ d\eta  \  \varphi(y) \ dy
\end{equation}
It has to reduce to the identity for $t=0$ and to represent the semiclassical approximation of the propagator. To this end,  the related phase function solves the H-J problem (\ref{eq-HJ-geom}) and moreover the  symbol  $b_0$ solves the regularized geometric version of the transport equation as in Theorem \ref{geo-tran2}. As we observed in Remark \ref{Remark-simb}, we require that in the region $\Upsilon_S$ free from critical points of $S$, the symbol $b_0$ behaves as $\lambda^{-n}(x,\eta) e^{-|\theta|^2}$.\\ 
Now, we easily see that
\begin{eqnarray}
\lefteqn{ \left(  H_x  - i \hbar \partial_t  \right)    e^{\frac{i}{\hbar}  S  }  b_{0}  } 
\nonumber\\
&=&   e^{\frac{i}{\hbar} S}   \left[ \hbar^0 \left(  \frac{|\nabla_x S|^2}{2m} + V(x) + \partial_t S   \right)    b_{0}  - i \hbar^1  \left(  \partial_t b_0 + \nabla_x S \ \nabla_x b_0  + \frac{\Delta_x S}{2m}    b_{0}  \right)  -     \frac{\hbar^2}{2m} \Delta_x  b_{0}   \right]   
\nonumber
\end{eqnarray}
The first two symbols of this sum vanish on the critical points set $\Sigma_S$ in such a way we can apply Corollary \ref{FIO-Y}, and so they realize bounded operators of order $O(\hbar^\infty)$. As  a consequence,
\begin{equation}
\label{eq-f0}
\left(  H_x  - i  \hbar  \partial_t  \right) \int_{\Bbb R^{n+k}}    e^{\frac{i}{\hbar}  S  }    b_{0}   d\theta \hat{\varphi}_\hbar (\eta) d\eta  =  -   \frac{\hbar^2}{2m}    \int_{\Bbb R^{n+k}}     e^{\frac{i}{\hbar} S   }    \Delta_x   b_{0}     d\theta \hat{\varphi}_\hbar (\eta) d\eta  + O(\hbar^\infty).
\end{equation}
The  operator $B_1(t)$ and the related symbol $b_1$ fulfill the analogous relation:
\begin{eqnarray}
\lefteqn{ \left(  H_x  - i \hbar \partial_t  \right)    e^{\frac{i}{\hbar}  S } \hbar b_{1}  } 
\nonumber\\
&=&   e^{\frac{i}{\hbar}S}   \left[ \hbar \left(  \frac{|\nabla_x S|^2}{2m} + V(x) + \partial_t S   \right)    b_{1}  - i \hbar^2  \left(  \partial_t b_1 + \nabla_x S \ \nabla_x b_1  + \frac{\Delta_x S}{2m}    b_{1}  \right)  -     \frac{\hbar^3}{2m} \Delta_x  b_{1}  \right]   
\nonumber
\end{eqnarray}
The transport equation  we now require for  this symbol is the following:
\begin{equation}
\label{trasp-01} \left\{
\begin{array}{l}
\displaystyle{ \partial_t b_1 + \nabla_x S \ \nabla_x b_1  + \frac{1}{2m} \Delta_x S  \ b_{1}   =   \frac{i}{2m}  \Delta_x  b_{0}   ,}
\\
\\
b_1(0,x,\eta,\theta) = 0.
\end{array}
\right.
\end{equation}
As  a consequence, 
\begin{eqnarray}
\label{eq-f56}
 \left(  H_x  - i  \hbar  \partial_t  \right) \int_{\Bbb R^{n+k}}     e^{\frac{i}{\hbar}  S  }   ( b_{0} + \hbar b_1 )   d\theta \hat{\varphi}_\hbar (\eta) d\eta   
=    - \frac{ \hbar^3 }{2m}  \int_{\Bbb R^{n+k}}   e^{\frac{i}{\hbar} S   }   \Delta_x   b_{1}   d\theta \hat{\varphi}_\hbar (\eta) d\eta  + O(\hbar^\infty)
\nonumber
\end{eqnarray}
The equation for the second order symbol:
\begin{equation}
\label{trasp-02} \left\{
\begin{array}{l}
\displaystyle{ \partial_t b_2 + \nabla_x S \ \nabla_x b_2  + \frac{1}{2m} \Delta_x S  \ b_{2}    =  \frac{i}{2m}  \Delta_x  b_{1},}
\\
\\
b_2(0,x,\eta,\theta) = 0.
\end{array}
\right.
\nonumber
\end{equation}
implies 
\begin{eqnarray}
\label{eq-f57}
\left(  H_x  - i  \hbar  \partial_t  \right) \int_{\Bbb R^{n+k}}    e^{\frac{i}{\hbar}  S  }   ( b_{0} + \hbar b_1 + \hbar^2 b_2 )   d\theta \hat{\varphi}_\hbar (\eta) d\eta  
=   - \frac{ \hbar^4 }{2m}  \int_{\Bbb R^{n+k}}   e^{\frac{i}{\hbar} S   }    \Delta_x   b_{2}     d\theta \hat{\varphi}_\hbar (\eta) d\eta  + O(\hbar^\infty)
\nonumber
\end{eqnarray}
Therefore we can deal with functions $b_j$,  $j \ge 1$ fulfilling the recurrent equations
\begin{equation}
\label{trasp-02} \left\{
\begin{array}{l}
\displaystyle{ \partial_t b_j + \nabla_x S \ \nabla_x b_j  + \frac{1}{2m} \Delta_x S  \ b_{j}   =   \frac{i}{2m}  \Delta_x  b_{j-1}  ,}
\\
\\
b_j (0,x,\eta,\theta) = 0.
\end{array}
\right.
\end{equation}
Each solution $b_j$ is a symbol as in Definition \ref{def-simb} and therefore (thanks to Th. \ref{bound-proof}) it defines a bounded operator $B_j(t)$. The same holds true  for the remainder operators:
$$
\mathcal{R}_j (t) \varphi =  (2\pi\hbar)^{-n} \int_{\Bbb R^{2n+k}}   e^{\frac{i}{\hbar} ( S( t,x,\eta,\theta)  - \langle y , \eta \rangle )   }   \hbar^{2+j} r_{j} ( t,x,\eta,\theta) \ d\theta  \ d\eta  \  \varphi(y) \ dy
$$ 
where $\ds r_j :=    \frac{i}{2m}  \Delta_x  b_{j}   $. In order to prove it, we need the following
\begin{lemma}
For all $j \ge 1$ the solution of equation (\ref{trasp-02}) fulfills the estimates:
\begin{equation}
|b_j|, |\Delta_x b_j|(t,x,\eta,\theta) \le  
\left\{
\begin{array}{l}
\displaystyle{
C_j^+ (T) e^{d_j (T) \lambda(x,\eta)}  e^{-|\theta|^2}, \quad (x,\eta,\theta) \notin \Upsilon_S  }
\\
\\
C_j^- (T) \lambda^{-n} (x,\eta) e^{-|\theta|^2}, \quad (x,\eta,\theta) \in \Upsilon_S.
\end{array}
\right.
\end{equation} 
\end{lemma}
\proof
We consider the following problem
\begin{equation}
\left\{
\begin{array}{l}
\displaystyle{ \frac{d}{d\tau}  \zeta(t,x,\eta,\theta)(\tau)  =  \nabla_x S (t, \zeta (t,x,\eta,\theta)(\tau), \eta, \theta )  }
\\
\\
\zeta(t,x,\eta,\theta)(t) = x
\end{array}
\right.
\end{equation}
and define 
$$
\Phi(\tau,x,\eta,\theta) :=   \exp  \left\{  -  \frac{1}{2m}  \int_0^\tau  \Delta_x S (t,\zeta(t,x,\eta,\theta)(r),\eta,\theta)  dr      \right\}
$$
in order to apply the theory of characteristics ($(\theta,\eta)$ fixed) and find the solution:
\begin{equation}
b_j (t,x,\eta,\theta) = \frac{i}{2m}  \int_0^t   \Phi(t-\tau,x,\eta,\theta)  \Delta_x  b_{j-1} (\tau,\zeta(t,x,\eta,\theta)(\tau),\eta,\theta)  d \tau
\label{rec-bj}
\end{equation}
By the iteration of this map, we have a direct linear relationship between $\Delta_x b_0$ and $b_j$. Now we recall the estimates on $b_0$ proved in Theorem \ref{Th-b_00}
$$
| \partial_x^\alpha b_0 (t,x,\eta, \theta )| \le  C_\alpha^+ (T) e^{d_{\alpha}(T) \lambda(x,\eta)}  e^{-|\theta|^2}  \quad \forall (x,\eta,\theta) \in \Bbb R^{2n+k}
$$
and the explicit analytic structure of $S$ studied in Theorem \ref{Rep-S}:
\begin{eqnarray}
S &=&  \langle x , \eta \rangle   - \frac{t}{2m} \eta^2 -  t \langle L x, x \rangle   +   \langle Q(t)  \theta  ,   \theta   \rangle 
+ \langle v(t,x,\eta) , \theta + f(t,x,\theta)  \rangle + \langle \nu(t,x,\theta) , \theta  \rangle 
\nonumber\\
&+&  g (t,x,\theta)
\end{eqnarray}
which implies the exponential behaviour of $\Phi$. The exponential upper bound for $|b_j|$ and $|\Delta_x b_j|$  follows directly from that. In the region $\Upsilon_S$ we recall the upper bound of type $\lambda^{-n} (x,\eta) e^{-|\theta|^2}$ we required for $\Delta_x b_0$ and the expansion $\Delta_x S (z) = \Delta_x S (\bar{z}) + G(z)$ with $G \in C^\infty_b$ (see Th. \ref{theo-S-eq}). By using (\ref{rec-bj}) we obtain this second estimate  also for $|b_j|$ and $|\Delta_x b_j|$.   
\endproof

As  a consequence, we can apply the boundedness result of  Theorem  \ref{bound-proof} and state the existence of constants  $K_j (T)>0$ such that  $\| \mathcal{R}_j (t) \| \le K_j (T) \hbar^{2+j}$. By well known  arguments related to the Duhamel formula we obtain the estimate:
$$
\Big\|  U(t) - \sum_{j=0}^N B_j (t)  \Big\| \le \frac{1}{\hbar} \int_0^t \|  \mathcal{R}_N (s)   \| \ ds \le T  K_N (T)  \hbar^{N+1}, \quad t \in [0,T]. 
$$ 
\endproof

Now we clarify the relationship between the construction of the previous theorem and Chazarain's formulation \cite{Chz}, as well as  with the integral representation of Fujiwara \cite{DF}.
\begin{theorem}
\label{th-corr86}
Let  $t\in[0,t_0]$,  with $t_0$ so small that the solution of the Hamilton-Jacobi equation does not develop caustics.  Consider the construction of Theorem 1.1, truncated at any finite order $J$:
\begin{equation}
\label{series-loc}
\sum_{j=0}^J  B_j(t) \varphi  :=  \sum_{j=0}^J  (2\pi\hbar)^{-n} \int_{\Bbb R^n} \int_{\Bbb R^n} \int_{\Bbb R^k}   e^{\frac{i}{\hbar} ( S( t,x,\eta,\theta)  - \langle y , \eta \rangle )   }   \hbar^j b_{j} ( t,x,\eta,\theta) \ d\theta  \ d\eta  \  \varphi(y) \ dy
\end{equation}
Then:
\begin{enumerate}
\item
\begin{equation}
\label{Ch-staz}
\sum_{j=0}^J  B_j(t) \varphi  =  \sum_{\alpha=0}^J  U^{ch}_\alpha (t) \varphi + O(\hbar^{J+1})
\end{equation}
Here $U^{ch}_\alpha(t)$ is the term of order $\hbar^\alpha$ of Chazarain's FIO (\cite{Chz}). 
\item
\begin{equation}
\label{Fu-staz}
\sum_{j=0}^J  B_j(t) \varphi  =  \sum_{\alpha=0}^J  U^{F}_\alpha(t) \varphi + O(\hbar^{J+1})
\end{equation}
where this time $U^{F}_\alpha(t)$ is the term of order $\hbar^\alpha$ of  Fujiwara's integral operator (\cite{DF}).
\end{enumerate}
\end{theorem}
\proof
In order to prove the first assertion,   the main idea is to apply the stationary phase theorem to the oscillatory integrals  (\ref{series-loc})  with respect to $\theta$-variables. In the same way, if we consider the stationarity argument with respect to $(\theta,\eta)$-variables we obtain the second assertion.\\
In the small time regime $t\in[0,t_0]$ there exists a unique smooth and global critical point $\theta^\star(t,x,\eta)$, solution of $0=\nabla_\theta S(t,x,\eta,\theta)$. This fact suggests us to consider the translated phase function around this point $S(t,x,\eta,\theta + \theta^\star(t,x,\eta))$ with $\theta \in B_1 (0)$ and symbol $b_0$ (see Theorem 1.1) for which we choose  the regularizing part as $\rho(\theta) := (vol B_1(0))^{-1} \mathcal{X}_1 (\theta)$, a $C^\infty$ cut off function for the ball $B_1(0)$. The compact behaviour of $b_j$ on the $\theta$-variables  follows as a consequence.  The uniqueness of $\theta^\star$ and the compact setting in the oscillatory integral allow us to apply the stationary phase theorem to each integral in the $\theta$-variables 
$$
B_j (t,x,\eta) = \int_{\Bbb R^k}   e^{ \frac{i}{\hbar}  S(t,x,\eta,\theta)    } \hbar^j   b_{j} ( t,x,\eta,\theta) \ d\theta 
$$
obtaining
\begin{equation}
\label{sim-corr}
B_j (t,x,\eta) =  e^{ \frac{i}{\hbar} S(t,x,\eta,\theta^\star)   }  |{\rm det} \nabla^2_\theta S(t,x,\eta,\theta^\star (t,x,\eta))|^{-\frac{1}{2}} \ e^{\frac{i\pi}{4}\sigma}  \hbar^j  b_{j} ( t,x,\eta,\theta^\star)   + O(\hbar^{j+1})
\end{equation}
where $\sigma= sgn \nabla^2_\theta S(t,x,\eta,\theta^\star (t,x,\eta))$ and we have omitted (to simplify the exposition) the explicit form of the higher orders symbols. Now we remark that the function   $S(t,x,\eta,\theta^\star)$ equals the phase used in the Chazarain's paper (the action functional evaluated on the classical curve with boundary conditions $x$ and $\eta$). Hence, by the uniqueness of the symbol expansion of the propagator in powers of $\hbar$ , we get the corrispondence between the symbols obtained  as in (\ref{sim-corr}) and the ones obtained in the above-mentioned paper.  This implies the equivalence of the two series (\ref{Ch-staz}) up to an order $o(\hbar^{J+1})$. By  the same argument, applied this time to  the integrals over $u:=(\theta,\eta)$ and $\Phi(t,x,y,u):=S( t,x,\eta,\theta)  - \langle y , \eta \rangle$   
$$
\widetilde{B}_j (t,x,y) = \int_{\Bbb R^n} \int_{\Bbb R^k}   e^{\frac{i}{\hbar} ( S( t,x,\eta,\theta)  - \langle y , \eta \rangle )   }   \hbar^j b_{j} ( t,x,\eta,\theta) \ d\theta  \ d\eta =  \int_{ \Bbb R^{n+k} }   e^{\frac{i}{\hbar}  \Phi(t,x,y,u)}  \hbar^j \widetilde{b}_{j} (t,x,u)     du
$$
we use the uniqueness of the critical point $u^\star (t,x,y)$. to get
$$
\widetilde{B}_j (t,x,y) =  e^{\frac{i}{\hbar}  \Phi(t,x,y,u^\star)     }   \hbar^j  |{\rm det} \nabla^2_\theta \Phi(t,x,y,u^\star (t,x,y))|^{-\frac{1}{2}} e^{\frac{i\pi}{4}\sigma} \   \widetilde{b}_{j}( t,x,u^\star(t,x,y))  + O(\hbar^{j+1})
$$
The phase function $\Phi( t,x,y,u^\star)$ is the same used by Fujiwara and therefore also  (\ref{Fu-staz}) is proved. This concludes the proof of the Theorem. 
\endproof

\noindent
\section{Multivalued WKB semiclassical approximation}
\renewcommand{\theequation}{\thesection.\arabic{equation}}
\setcounter{equation}{0}%
In this final section we prove Theorem 1.2, mainly applying the Stationary Phase theorem to the global FIO  (\ref{def-semi}), in order to get a multivalued WKB semiclassical approximation of the Schr\"odinger evolution operator.

\noindent
{\bf Proof of Theorem 1.2}\\
We start by recalling that (as proved in Theorem 1.1) the $\hbar$-Fourier Integral Operator
\begin{equation}
\label{semi-B0}
B_0(t) \varphi := (2\pi\hbar)^{-n} \int_{\Bbb R^n} \int_{\Bbb R^n} \int_{\Bbb R^k}   e^{\frac{i}{\hbar} ( S( t,x,\eta,\theta)  - \langle y , \eta \rangle )   }  b_{0} ( t,x,\eta,\theta) \ d\theta  \ d\eta  \  \varphi(y) \ dy
\end{equation}
is a semiclassical approximation of the Schr\"odinger propagator for all $t \in [0,T]$.
Under the particular hypothesis 
$$
V(x) = \frac{1}{2}|x|^2 + V_0 (x), \quad \sup_{x\in \Bbb R^n} \| \nabla^2 V_0 (x) \|<1, \quad  t \neq (2\tau +1)\frac{\pi}{2} \ (\tau\in \mathbb{N}),
$$
we proved (see Theorems \ref{isol}, \ref{Th-locS} and \ref{th-grad11}) that the phase function has isolated and finitely many critical points; precisely the equation
\begin{eqnarray}
\label{staz99}
\nabla_\theta S(t,x,\eta,\theta) = 0
\end{eqnarray}
is solved on a finite open partition $\ds (x,\eta) \in \Bbb R^{2n} = \bigcup_{\ell=1}^{\mathcal{N}(t)} D_\ell$ in such a way that  on each $D_\ell$ there are exactly $\ell$ smooth solutions $\theta^\star(t,x,\eta)$, $1\le  \alpha \le \ell$. This property allows us to apply the Stationary Phase Theorem (see \cite{Ho2} vol. I) to the oscillatory integral in (\ref{semi-B0}). The result is:
\begin{eqnarray}
B_0 (t,x,\eta)\Big|_{D_\ell}&=&\int_{\Bbb R^k}   e^{  \frac{i}{\hbar}  S(t,x,\eta,\theta)   }  b_{0} (t,x,\eta,\theta) \ d\theta
\nonumber\\
&=&\sum_{\alpha=1}^\ell    e^{\frac{i}{\hbar} S ( t,x,\eta,\theta^\star_\alpha (t,x,\eta))   }     |{\rm det} \nabla^2_\theta S(t,x,\eta,\theta^\star_\alpha (t,x,\eta))|^{-\frac{1}{2}}  e^{\frac{i\pi}{4}\sigma_\alpha} b_{0} (t,x,\eta,\theta^\star_\alpha (t,x,\eta))  
\nonumber\\
&+& O(\hbar)
\nonumber
\end{eqnarray}
where $\sigma_\alpha = sgn \nabla^2_\theta S ( t,x,\eta,\theta^\star_\alpha (t,x,\eta)) $.
In the small time regime $t \in [0,t_0]$ and for  potentials $V$ quadratic at infinity, it is well known (see i.e. \cite{W1}) that the graph of the Hamiltonian flow 
\begin{eqnarray*}
\Lambda_t &:=& \left\{  (y,\eta;x,p) \in T^\star \Bbb R^n \times T^\star \Bbb R^n \ | \ (x,p) = \phi_\H^t (y,\eta)     \right\}
\nonumber\\
&=& \left\{ (y,\eta;x,p) \in T^\star \Bbb R^n \times T^\star \Bbb R^n \ | \  p = \nabla_x S, \; y =  \nabla_\eta S, \;  0 =  \nabla_\theta S \right\}  
\end{eqnarray*}
is globally transverse to the base manifold $(x,\eta) \in \Bbb R^{2n}$, so the equation  (\ref{staz99})  admits a unique global smooth solution $\theta^\star (t,x,\eta)$.  This simplified setting yields:
\begin{eqnarray}
B_0 (t,x,\eta) &=&  e^{\frac{i}{\hbar} S ( t,x,\eta,\theta^\star (t,x,\eta))   }  \   |{\rm det} \nabla^2_\theta S(t,x,\eta,\theta^\star (t,x,\eta))|^{-\frac{1}{2}} \ e^{\frac{i\pi}{4}\sigma} \ b_{0} (t,x,\eta,\theta^\star (t,x,\eta))  
\nonumber\\
&+& O(\hbar)
\nonumber
\end{eqnarray}
which is the usual WKB construction, local in time.
\endproof
    .


\end{document}